\documentclass[aps,pre,twocolumn,showpacs]{revtex4}

\usepackage{epsfig}
\usepackage{graphicx}
\usepackage{amsmath}
\usepackage{amssymb}
\usepackage{amsfonts}
\usepackage{dcolumn}
\draft

\begin{document}

\title{Mixed symmetry localized modes and breathers in binary mixtures of Bose-Einstein condensates in optical lattices}
\author{H. A. Cruz$^{1,2}$, V. A. Brazhnyi$^1$, V. V. Konotop$^{1,2,3}$, G. L. Alfimov$^4$, and M. Salerno$^5$}

\address{$^{1}$Centro de F\'{\i}sica Te\'orica e Computacional,
Universidade de Lisboa, Complexo Interdisciplinar, Avenida
Professor Gama Pinto 2, Lisboa 1649-003, Portugal
\\
$^2$Departamento de F\'{\i}sica,
Universidade de Lisboa, Campo Grande, Ed. C8, Piso 6, Lisboa
1749-016, Portugal
\\
$^3$Departamento de Matem\'aticas, E. T. S. Ingenieros Industriales,
Universidad de Castilla-La Mancha, 13071 Ciudad Real,
Spain
\\
$^4$Moscow Institute of Electronic Engineering,
    Zelenograd, Moscow, 124498, Russia
\\
$^5$Dipartimento di Fisica "E. R. Caianiello", Consorzio Nazionale
Interuniversitario per le Scienze Fisiche della Materia (CNISM),
Universit\'{a} di Salerno, I-84081, Baronissi (SA), Italy}

\begin{abstract}
\par
We study localized modes in binary mixtures of Bose-Einstein
condensates embedded in one-dimensional optical lattices.  We
report a diversity of asymmetric modes and investigate their dynamics.
We concentrate on the cases where one of the components is
dominant, i.e. has much larger number of atoms than the other one,
and where both components have the numbers of atoms of the same
order but different symmetries. In the first case we propose a
method of systematic obtaining the modes, considering the "small"
component as bifurcating from the continuum spectrum. A
generalization of this approach combined with the use of the
symmetry of the coupled Gross-Pitaevskii equations allows
obtaining breather modes, which are also presented.
\end{abstract}
\pacs{...}

\maketitle

\section{Introduction}

Optical lattices are known to be an effective tool for
manipulation of matter waves. The induced spatial periodicity
allows, in particular, for the existence of various localized
structures, not available in homogeneous systems. Such
excitations, which are called localized modes (also gap solitons
depending on the limit they are considered in) have been
intensively studied during the last few years (see e.g. recent
reviews \cite{olI,olII} and references therein) and observed
experimentally in  a Bose-Einstein condensate (BEC) of $^{87}$Rb
atoms~\cite{GS_exp}. While main attention was devoted to
single-component BECs, existence of localized modes was found to
be a generic property of multicomponent systems, as well. More
specifically, emergence of coupled bright solitons, from the
modulational instability of binary mixtures of BECs in optical
lattices (OLs), was observed numerically in \cite{2comp_periodic}. In
\cite{2comp_localization}, it was shown that OLs can
be used to modify the interactions both within and between BEC
components, leading to the creation of two-component gap solitons.
Existence and stability of 2D and 1D coupled gap solitons in a
binary BEC with repulsive interactions was analyzed by means of
the variational approximation and numerically in
\cite{2comp_1d2d}. It was also found that localized modes can
exist in atomic-molecular condensates~\cite{ambec}, in
Bose-Fermi mixtures in OL~\cite{msBFM,bose-fermion} 
and in three-component spinor condensates~\cite{ostr-kiv}.

All the results on binary mixtures of BECs reported so far, concern
excitations of the two components characterized by similar
scales, i.e. by spatial localization regions, and numbers of
particles of the different species having the same order. Also
earlier studies were restricted to modes having density
distributions in both components of the same symmetry, whenever
the components belong to the same gap (gap solitons belonging to
different gaps were addressed in~\cite{2comp_1d2d}).

In the present paper we report a diversity of novel states
observable in two-component BECs embedded in OLs with
different  symmetry properties of the density distributions with
respect to the periodic potential and therefore termed below as
{\em mixed symmetry} modes (or also gap solitons of mixed
symmetry).  The interaction between the two components of the
condensate is crucial for the existence of these modes since they
are formed from single component modes which have different
stability properties (one stable and the other unstable) in
absence of interaction. The interspecies interaction allows for
stabilization of the unstable mode in the pair and subsequently
for formation of a stable bound state. Mixed symmetry modes exist
both for attractive and repulsive inter-- and intra--species
interactions and represent very stable new excitations of two
component BECs in OLs. Stabilization phenomena
induced by the interspecies interaction are also possible for
equal symmetry states (both symmetric or antisymmetric with
respect to a minimum or a maximum of the potential). In this case
we find that modes which are both unstable in absence of
interaction can form stable equal symmetry bound states in
presence of interspecies interaction. Besides stationary states of
mixed symmetry  we show that non stationary states of mixed
symmetry which give rise to breather-like oscillations are also
possible. When the inter-- and the intra--species interactions are
all equal, the two-component breather modes can be constructed
from a linear superposition of two mixed symmetry modes with an
unbalanced  number of atoms in the two components. On the other
hand, we show that breather modes with a balanced numbers of atoms
and with a very regular dynamics which persist on a long time
scale without any apparent emission of matter, are also possible.
Mixed symmetry modes, both of stationary type and of breather
type, considerably  enlarge the number of stable localized
excitations which one can find in BEC mixtures, this showing the
richness of these systems with respect to single component BECs.

The paper is organized as follows. In Sec. \ref{sec_model} we recall
the system of coupled Gross-Pitaevskii (GP) equations describing the
mixture, specify the parameters to be used for particular numerical
simulations, and give a simple physical picture suggesting existence of
asymmetric  modes.  In Sec. \ref{sec_numerics} we discuss on
particular examples the localized modes of mixed symmetry in the limit
($N_1\gg N_2$) and study stability of the modes by direct numerical
simulations. In Sec.~\ref{LargeAmpl} we consider large amplitude
localized modes both of equal symmetry and of mixed symmetry type.
Sec.~\ref{breathers} is devoted to breathing modes, both with balanced
and unbalanced number of atoms, which are obtained by combining the
developed approach with the symmetry of the coupled GP equations. In
the Conclusion the main outcomes are summarized.

\section{Statement of the problem}
\label{sec_model}

\subsection{The model}

At the zero temperature and a large number of atoms, when quantum
fluctuations can be neglected, a diluted binary mixtures of
BECs is well described by the coupled GP equations
($j=1,2$)~\cite{bec_book}
\begin{eqnarray}
&&i\hbar\frac{\partial\Psi_{j}}{\partial t} = -\frac{\hbar^2}{2m_{j}}\Delta \Psi_{j} + V_{j}({\bf r})\Psi_{j}
\nonumber \\
&&\quad +\frac{2\pi\hbar^2}{M} \Bigg(\frac{2M}{m_j}a_{jj}|\Psi_{j}|^2 +a_{j,3-j}|\Psi_{3-j}|^2 \Bigg) \Psi_{j}.
\label{NLS_GP}
\end{eqnarray}
Here $m_j$ is the atomic mass of the $j^{th}$ component,
$M=m_1m_2/(m_1+m_2)$ is the reduced mass, and $a_{ij}=a_{ji}$ are
the s-wave scattering lengths of the binary interactions.
The macroscopic wave functions, $\Psi_{j}$,  are normalized to
numbers of atoms $N_j$: $\int|\Psi_{j}|^2d{\bf r}=N_j$, which do
not depend on time. The trap potential for the component $j$,
$V_j({\bf r})$, consists  of a superposition of a magnetic trap
and an OL along the $x$-direction:
\begin{eqnarray}
\label{Vj}
V_{j}({\bf r}) = \frac{m_j\omega_{j}^2}{2} ({\bf r}_{\perp}^2+ \lambda^2 x^2) + V_{0}E_{Rj}U(\kappa x).
\end{eqnarray}
Here $\omega_{j}$ is the transverse linear oscillator frequency of
the component $j$,  $\lambda$ is the aspect ratio (it is assumed
to be equal for both components), $d$ is the lattice constant,
$\kappa=\pi/d$, $V_0$ is the amplitude of OL measured
in terms of the respective recoil energy $E_{Rj}=\hbar^2
\kappa^2/(2m_j)$, and $U(\cdot)$ is a $\pi$-periodic dimensionless
function (the lattice profiles are considered to be the same
for both components).

We consider cigar-shaped condensates, i.e. the limit $\lambda\ll
1$, at low densities. To specify the scaling we introduce a small
parameter of the problem  $\epsilon=\kappa a$, where $a$ is an
average transverse oscillator length defined as
$a=\sqrt{\hbar/(2M\omega)}$, where
$\omega=\sqrt{\omega_1\omega_2}$. Then the applicability of the
theory is determined as smallness of a number of particles on
the scale of the scattering length, i.e. as
$\sqrt{n|a_{11}|}\sim\kappa a=\epsilon \ll 1$, where
$n=N\sqrt{\lambda}/a$ is the linear atomic density and $N=N_1+N_2$
is the total number of atoms.   This is the case when the
multiple-scale analysis can be applied to reduce (\ref{NLS_GP}) to
a system of coupled 1D GP equations (see
e.g.~\cite{2comp_periodic,olI}). By introducing slow dimensionless
independent variables $X=\epsilon x/a$ and $T=\epsilon^2 \omega t$
and dimensionless  order parameter $\Phi_{j}= {\cal O}(1)$,
through the expansion  ($j=1,2$)
\begin{eqnarray}
\label{solution}
\Psi_{j} = \frac{\kappa}{\sqrt{ 2|a_{11}|} }
\left(\Phi_{j}+ \epsilon\Phi_{j}^{(1)} + ...\right)\varphi_j({\bf R}_{\perp})e^{-i \omega_{j} t}
\end{eqnarray}
with the normalized transverse eigenfunctions
\begin{eqnarray*}
\varphi_j ({\bf R}_{\perp}) = \sqrt{\frac{m_j}{2\pi
M}}\left(\frac{\omega_j}{\omega_{3-j}}\right)^{\frac14} \exp\left(
-\frac{m_j}{4M}\sqrt{\frac{\omega_j}{\omega_{3-j}}}{\bf
R}_{\perp}^2\right),
\end{eqnarray*}
we obtain a system of coupled 1D GP equations
\begin{eqnarray}
\label{GP_1D_1}
i\frac{\partial\Phi_{j}}{\partial T} = -\frac{M} {m_j}
\frac{\partial^2 \Phi_{j}}{\partial X^2} -  V_{0j}U(X)\Phi_{j}
\nonumber \\
  + (\chi_{j}|\Phi_{j}|^2 + \chi|\Phi_{3-j}|^2) \Phi_{j}\,.
\end{eqnarray}
Here $V_{0j}=V_0{M}/{m_j}$ and the nonlinear coefficients have the following form
\begin{eqnarray*}
\chi_{j}=\sqrt{\frac{\omega_j}{\omega_{3-j}}}\frac{a_{jj}}{|a_{11}|}
\quad \mbox{and}\quad
\chi =  \frac{(m_1+m_2)\sqrt{\omega_1\omega_2}}{(m_1\omega_1+m_2\omega_2)}\frac{a_{12}}{|a_{11}|}. 
\end{eqnarray*}


Looking for stationary solutions $\Phi_{j} \rightarrow \Phi_{j}e^{-i\mu_jT}$, we obtain the system
\begin{eqnarray}
\label{eigen}
\mu_{j}\Phi_{j} = -\frac{M}{m_j} \frac{d^2 \Phi_{j}}{d X^2}  &-& V_{0j}U(X)\Phi_{j} \nonumber \\
&+& \left(\chi_{j}|\Phi_{j}|^2 + \chi|\Phi_{3-j}|^2 \right)\Phi_{j}
\end{eqnarray}
with $\mu_{1}$ and $\mu_{2}$ being the chemical potentials of the first
and second components,  respectively.

\subsection{The main goal and the terminology}

The goal of this paper is to present some families of localized
modes  and breathers of Eqs.~(\ref{eigen}) which are typical
for multicomponent BECs and which posses interesting symmetry
properties. These solutions have been found numerically by means
of different methods. The typical difficulty which arise in this
problem is to get {\it a priori} idea on possible shapes of
localized modes which can be used for numerical iterative
algorithm.  To overcome this difficulty we have used two
alternative approaches. The first involves a search of
bifurcations for specific solutions of (\ref{eigen}) using
shooting method (see \cite{AKS}) as the main tool. The second is
the self-consistent approach which has been used before for
applications to single-component BECs (see \cite{LP05}).  The self
consistent method, indeed, is very convenient for exploring the
band structure in presence of nonlinearity and to address to
specific modes for which good initial guesses are available.
Basically, the shooting method allows one to make a systematic
exploration of the whole families of solutions in terms of a
single (for single component BECs) parameter (shooting parameter).
Similar exploration of the case of coupled GP equations is quite
complicated, since in this case one has to scan a semi-infinite
domain of two shooting parameters. In the present paper we use the
shooting method to investigate localized modes of the mixed
symmetry type in the limit $N_1\gg N_2$ (see Sec.
\ref{sec_numerics}) and resort to self consistent approach to
investigate BEC mixtures with balanced number of atoms (see Sec.
\ref{LargeAmpl}). When one of the components is dominant, i.e. it
has much larger number of atoms than the second component, the
last one can be considered as bifurcating from the linear
spectrum. Such modes will be referred to as {\it unbalanced}
or strongly asymmetric in number of atoms. In this way after
constructing the localized solution for the first component in the
absence of the second one, the mode can be constructed by means of
some iterative procedure. Combining this approach with the
symmetry of the system we will also be able to construct
(numerically) exact breathing localized modes.

Another regime we will deal with corresponds to the situation when
the both components are large amplitude solutions having numbers
of particles of the same order, $N_1\sim N_2$. In this case we
will show existence of not only simplest fundamental modes with
equal symmetry but also modes and breathers with mixed symmetry.
We remark that mixed symmetry modes were also found in
atomic-molecular condensates~\cite{ambec}. In this case, however,
the mean-field equations had an intrinsic asymmetry due to the
different properties of atoms and molecules.

First of all it is desirable to establish a
convenient terminology for the identification of modes. To this end we
observe that multiplying (\ref{eigen}) by $\partial
\Phi_j / \partial X$ and integrating with respect to $X$  one arrives at
the {\em necessary condition} for existence of a stationary
localized mode ($\Phi_{1},\Phi_{2}$):
\begin{eqnarray}
    \label{necessary}
    \int\left(V_{01}|\Phi_{1}|^2 +V_{02}|\Phi_{2}|^2\right)\frac{dU(X)}{dX}dX=0.
\end{eqnarray}
For the physical interpretation of this condition, we notice that a solution of (\ref{eigen}) minimizes the Hamiltonian 
$ 
H=H_1+H_2+H_{12}+H_{v}(0),
$ 
 where
\begin{eqnarray}
    \label{hamilt}
     && H_j=\int\left(\frac{M}{m_j}\left|\frac{\partial
\Phi_{j}}{\partial
X}\right|^2+\frac{\chi_j}{2}|\Phi_{j}|^4\right)dX
\end{eqnarray}
$ H_{12}=\chi\int  |\Phi_{1}|^2|\Phi_{2}|^2dX$, and we have defined
\begin{eqnarray}
    H_v(\zeta)\equiv\int\left(V_{01}|\Phi_{1}|^2 +V_{02}|\Phi_{2}|^2\right)U(X-\zeta)dX.
\end{eqnarray}
Considering $(\Phi_1(X-\zeta),\Phi_2(X-\zeta))$ with $\zeta\ll 1$, which for a stable solution must
lead to increase of the energy, we immediately obtain  
\begin{eqnarray}
\label{h-deriv}
\frac{dH_v(\zeta)}{d\zeta}\Bigg|_{\zeta=0}=0\quad \mbox{and}\quad \frac{d^2H_v(\zeta)}{d\zeta^2}\Bigg|_{\zeta=0}>0.
\end{eqnarray}
The first of the obtained constrains is exactly Eq.~(\ref{necessary}).

Let us now require the potential to have a definite symmetry, i.e.
to be either even or odd, and concentrate on the solutions
possessing a specific symmetry, i.e.
$
\Phi_j(-X)=\pm  \Phi_j(X).
$
If the potential is odd, $U(-X)=- U(X)$, such 
solutions strongly localized about $X=0$ cannot satisfy
(\ref{necessary}). This allows us to conjecture nonexistence of
stable solitons localized about the point $X=0$ for odd
potentials. In particular, in the case of cos-like potential
\begin{eqnarray}
\label{cos}
    U(X)=2\cos(2X)\,,
\end{eqnarray}
explored below in the present paper: localized modes with a
defined symmetry can be excited only in the vicinity of one of the
points $X_n=\pi n$ where $n$ is an integer.

Considering solutions localized about $X=0$ for an even potential $U(X)=U(-X)$ one can
identify four types of the modes. They can be classified in
terms of the symmetry properties acquired with respect to the
periodic potential, considered earlier for a atomic-molecular
BEC~\cite{ambec} and for a single-component BEC~\cite{LP05}.

The diversity of modes makes it is desirable to establish a
convenient terminology for their identification. To this end we
classify states according to their symmetry properties with
respect to the minima of the periodic potential around which are
localized. A classification of this type was introduced in Refs.
\cite{ambec,LP05}, for atomic-molecular BECs and
single-component BECs, respectively, in terms of four single
component  states. These states are the OS -- {\em on-site
symmetric} and OA -- {\em on-site antisymmetric} modes, which are
respectively symmetric and antisymmetric with respect to a minimum
of the OL, hereafter referred as {\em lattice site}, and the IS --
{\em inter-site symmetric} and IA -- {\em inter-site
anti-symmetric} modes  which are respectively symmetric and
antisymmetric with respect to a maximum of the OL (the center of
the symmetry is the middle point between two lattice sites).
Notice that these symmetry properties are the same as for
intrinsic localized modes of nonlinear lattices, as one could
expect from the tight-binding approximation based on the
properties of Wannier functions~\cite{Kohn}  and reducing the
periodic GP equations to discrete lattices~\cite{AKKS,ambec}.
Except for the OS solution, which is always stable, the other
types of gap solitons are usually unstable in the one component
case~\cite{LP05}. However, from (\ref{h-deriv}) one can expect
that gap solitons which have the same symmetry properties in both
components should exist and may be stable. The respective modes
will be characterized by a double indexing, like for example
OS-OS, OS-OA, etc., the first and second pairs of letters
referring to the first and second components, respectively. The
modes OS-OS, OA-OA, IS-IS, and IA-IA will be called the {\em
"equal-symmetry" two-component modes}, while modes OS-OA, IS-IA,
etc. will be called mixed symmetry modes, with an evident meaning
in the notation.

\subsection{Spinor condensates}

Although our results have general character, and can be applied to
any binary mixture of BECs, either spinor or of different species,
to be specific we focus on spinor BECs. Such condensates are
available experimentally. As example we mention mixtures of the
hyperfine states  $|F=1, m_{F}=-1\rangle$ and $|F=2,
m_{F}=1\rangle$ as well as of the states $|F=2, m_{F}=1\rangle$
and $|F=2, m_{F}=2\rangle$ of $^{87}$Rb atoms produced in
\cite{2comp_exp_spinora} and \cite{2comp_exp_spinorw},
respectively. For a spinor mixture $m_1=m_2=m$, $M=m/2$, and the
potential (\ref{cos})  the  equations (\ref{eigen}) are simplified
\begin{subequations}
\label{eigen_spinor}
\begin{eqnarray}
\label{eigen1_spinor}
\mu_{1}\Phi_{1} = -\frac{1}{2} \frac{\partial^2 \Phi_{1}}{\partial X^2}
&-&V_{0}\cos(2X)\Phi_{1} \nonumber \\
&+& \left(\chi_{1}|\Phi_{1}|^2 + \chi|\Phi_{2}|^2 \right)\Phi_{1}\\
\label{eigen2_spinor}
\mu_{2}\Phi_{2} = -\frac{1}{2} \frac{\partial^2 \Phi_{2}}{\partial X^2}
&-&V_{0}\cos(2X) \Phi_{2}\nonumber \\
&+& \left(\chi|\Phi_{1}|^2 + \chi_{2}|\Phi_{2}|^2 \right)\Phi_{2}
\end{eqnarray}
\end{subequations}
where nonlinear coefficients are given by
$\chi_{1}=\Omega^{-1}a_{11}/|a_{11}|$, $\chi_{2}=\Omega
a_{22}/|a_{11}|$, and $\chi =
2a_{12}\Omega/[|a_{11}|(\Omega^2+1)]$ with
$\Omega=\sqrt{{\omega_2}/{\omega_1}}$.

In order to fix parameters we recall that the thermodynamical
stability of a homogeneous binary mixture is determined by
the condition~\cite{Pethick} $\Delta>0$, where
\begin{eqnarray}
\Delta=\chi_{1}\chi_{2}-\chi^2 \,.
\label{stab_cond}
\end{eqnarray}
Using this fact as the reference point we will consider the following
three situations $\Delta>0$, $\Delta<0$ and $\Delta=0$
(we however emphasize that, generally speaking, these condition should
not be considered as the stability/instability conditions for localized modes).
Before going
into details of numerical simulations, we observe that interchange of
the regimes with different $\Delta$ can be experimentally achieved
either by means of Feshbach  resonance affecting scattering lengths
$a_{ij}$ or by means of using transverse magnetic trap, i.e. geometry
of the trap,  resulting in different $\omega_j$ for different
components, say, in $\Omega=\sqrt{2}$ for the mixtures of hyperfine
states $|F=2, m_{F}=1\rangle$ and $|F=2, m_{F}=2\rangle$ of $^{87}$Rb
atoms  in a magnetic trap~\cite{2comp_exp_spinorw}.

\section{Localized modes of repulsive BEC mixtures with
unbalanced number of atoms (limit $N_1\gg
N_2$).} \label{sec_numerics}

\subsection{Defect modes}

\begin{figure}[ht]
\epsfig{file=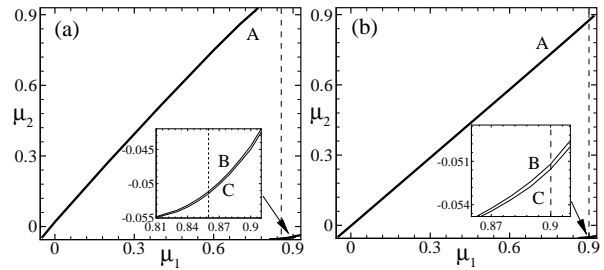,width=8cm} \caption{Discrete levels in the
first lowest gaps $\mu_{1,2}\in[-0.055,0.92]$. The insets show
zoom of branches B (OA symmetry) and C (OS symmetry). The
amplitude of the OL is $V_{0}=1$. (a) and (b)
correspond to different sets of nonlinear coefficients
$\chi_1$:$\chi_2$:$\chi$ such as in (a) $1$:$1.584$:$1.231$
($\Delta>0$) and in (b) $1$:$0.942$:$0.971$ ($\Delta<0$, these
relations are taken from ~\cite{2comp_exp_spinora}). The dashed
lines in (a) and (b) correspond to the cross sections at
$\mu_1=0.86$ and $\mu_1=0.9$, respectively.} \label{fig_w1w2}
\end{figure}

Let us consider the limit $N_2\to 0$. For the first step, neglecting
contribution of the second component, $|\Phi_2|^2$,  one can solve
equation (\ref{eigen1_spinor}) in the form
\begin{eqnarray}
\label{mode1} \mu_{1}\Phi_{1}^{(0)} = -\frac{1}{2}
\frac{\partial^2 \Phi_{1}^{(0)}}{\partial X^2}
-V_0\cos(2X)\Phi_{1}^{(0)} + \chi_{1}|\Phi_{1}^{(0)}|^2
\Phi_{1}^{(0)}
\end{eqnarray}
to find localized modes for a given value of the chemical
potential $\mu_1$. In what follows we focus only on symmetric
modes belonging to the lowest branch of the solutions, and
therefore referred to as the {\em fundamental mode}.
 As the second step, in the same approximation,
i.e. neglecting $|\Phi_{2}|^2 $, one can consider
(\ref{eigen2_spinor}) with $|\Phi_1^{(0)}|^2$ found from
(\ref{mode1}):
\begin{eqnarray}
\label{mode2_eigen} \mu_{2}\Phi_{2}^{(0)} = -\frac{1}{2}
\frac{\partial^2 \Phi_{2}^{(0)}}{\partial X^2} +
V_{eff}(X)\Phi_{2}^{(0)}
\end{eqnarray}
where the effective potential is defined as:
\begin{eqnarray}
\label{Veff} V_{eff}(X)=-V_0\cos(2X) + \chi|\Phi_{1}^{(0)}|^2.
\end{eqnarray}
After finding  discrete eigenvalues $\mu_2$ for a given $\mu_1$, one
can scan the gap for the first component varying $\mu_1$ and obtaining
all eigenvalues $\mu_2(\mu_1)$. In Fig.~\ref{fig_w1w2} we present the
results of this procedure for the cases $\Delta>0$ and $\Delta<0$.

The physical meaning of the above procedure consists in the
approximation of a localized modes $\Phi_1$, $\Phi_2$ in the limit
$N_2\to 0$. In this limit, the  wavefunction of the second
component, $\Phi_2$, appears as a {\em defect mode} and, strictly
speaking,  does not describe the atomic distributions at a finite
$N_2$. Since however, the solution at $N_2\approx 0$ is known,
departing from it and using an iterative procedure one can
construct the solutions for $N_2>0$. This approach is based on the following important statement (whose arguments are presented in Appendix~\ref{app:1}): 
{\it If $\mu_1$ and $\mu_2$ as well as
$V_{01}$ and $V_{02}$ are fixed, then the localized modes of
(\ref{eigen}) are isolated}. 
When $\mu_{1,2}$ varies one can speak
about {\it families} of the nonlinear localized modes. 

Before performing the search of localized modes, let us discuss in more details,
the results of the above analysis of defect modes. 
In Fig.~\ref{fig_w1w2} three branches of the solutions are shown for
different signs of $\Delta$. The branches A correspond to the case
where the chemical potentials are approximately equal. In order to
understand their occurrence let us consider a solution of
Eqs.~(\ref{eigen_spinor}) in the form $\Phi_{2} = \alpha\Phi_{1}$ where
$\alpha$ is a constant, which without loss of generality can be
considered real. Such a solution exists for  $\mu_1=\mu_2$ and
$\alpha^2=(\chi-\chi_1)/(\chi-\chi_2)>0$ and below will be referred to
as a {\em trivial solution}. In order to obtain the linear limit
(\ref{mode1}), (\ref{mode2_eigen}) one has to consider $\alpha\to 0$,
what is possible only for $\chi_1=\chi$. In this last case the branches
A would coincide with the diagonals of the boxes in
Fig.~\ref{fig_w1w2}. In the cases at hand $\chi_1\approx\chi$, what
explains deviations of the lines A from the diagonals.

\begin{figure}[ht]
\epsfig{file=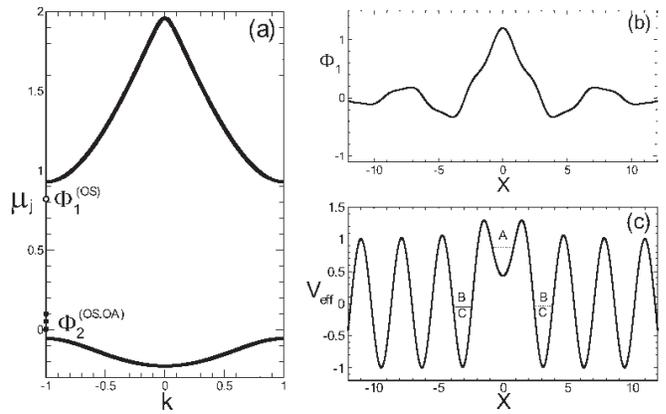,width=\columnwidth}
\caption{
(a) Band structure and localized levels for the values of parameters fixed as the case $\Delta>0$ in
Fig.~\ref{fig_prof}: the fundamental component at $\mu_1=0.86$ (open circle) has OS profile,
while the second component (filled circles) is given for both OS and OA profiles
considered at the same values of $\mu_2=0; 0.05, 0.1$ (see corresponding profiles in
Fig.~\ref{fig_prof}).
(b) The profile of the fundamental mode, $\Phi_1$, obtained in the limit $N_2=0$.
(c) The shape of the respective effective potential (\ref{Veff}).
}
\label{fig_Veff}
\end{figure}

The most interesting results, however, are shown in the insets of
Fig.~\ref{fig_w1w2}, where one can observe the two other branches of
the solution of the eigenvalue problem (\ref{mode2_eigen}) -- branches
B and C. These modes appear in the vicinity of the opposite gap edges
for the first and the second components, as this is illustrated in
Fig.~\ref{fig_Veff}(a) for the case $\Delta>0$: the eigenvalues are
located near the top of the gap for the first component and in the
vicinity of the bottom of the gap for the second component. For $N_2>0$
(but small) these eigenvalues give rise to OS and OA modes which are
characterized by a small difference in energies. Their shapes are shown
in Fig~\ref{fig_prof}.

As we have already mentioned these solutions appear as defect
modes, what is illustrated in  Fig.~\ref{fig_Veff} where we show
the fundamental localized mode of the first component near the
upper edge of the  gap (panel (b)), corresponding potential
$V_{eff}$ and the respective energy levels, which can be
interpreted as defect modes (panel (c)). The two eigenvalues B and
C  can be viewed as a splitting of discrete defect level (similar
to the level splitting in a double-well potential).

A number of the defect modes depends on the amplitude of the potential
$V_0$.  In all our cases the energy differences between the 
modes were very small, what makes it  convenient to speak about appearance of the pairs of the modes.
 Then, one can
determine $V_n$, such that for $V_n<V_0<V_{n+1}$ there exist $n$ pairs
of the defect modes (simultaneously with the fundamental mode).
In particular, for the cos-like potential and $\Delta>0$ ($\Delta<0$) we have found numerically that
$V_1\approx 0.43$, $V_2\approx 1.42$ ($V_1\approx 0.61$, $V_2\approx 2.01$).


\subsection{The modes with   $N_1\gg
N_2$.}

\begin{figure}[ht]
\epsfig{file=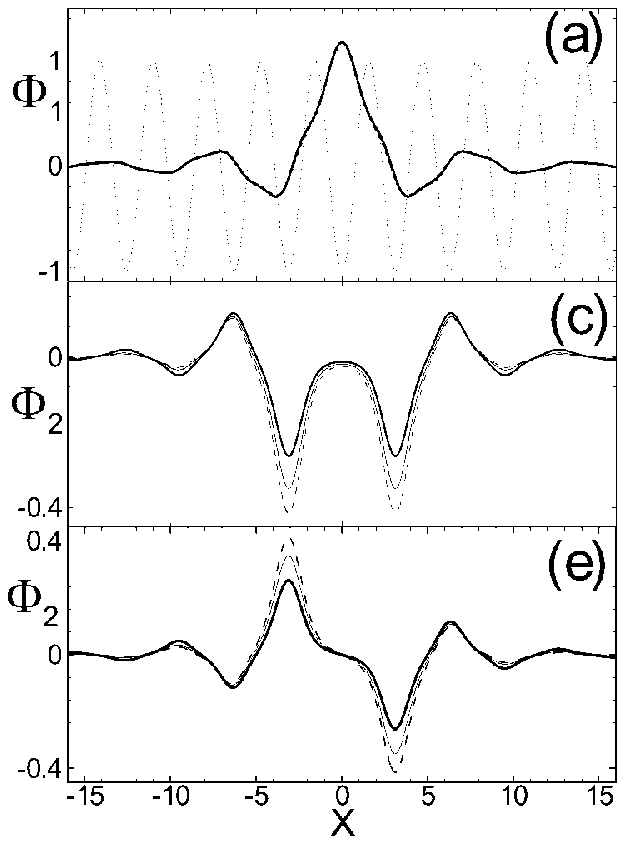,width=4cm}\epsfig{file=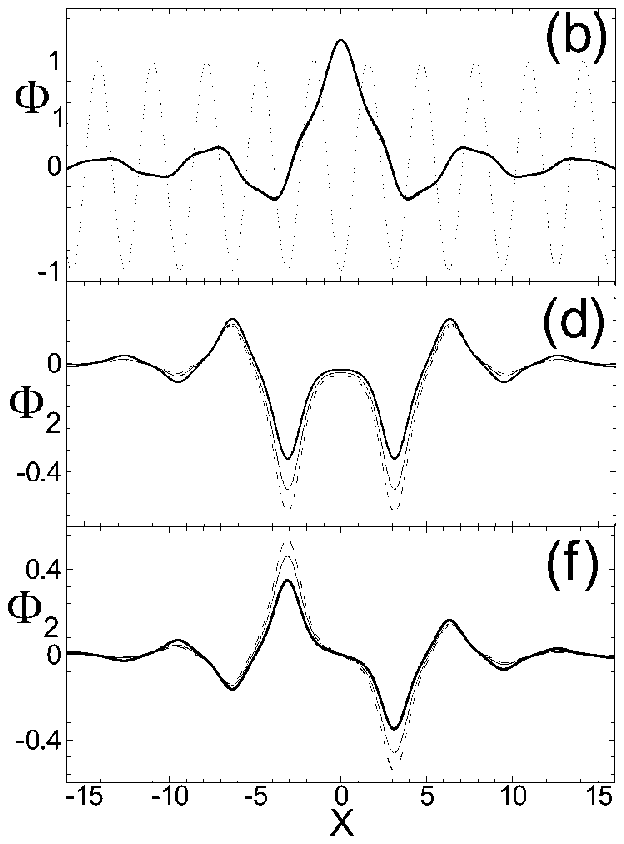,width=4cm}
\caption{Localized OS (panels (a)-(d)) and  OA (panels
(e) and (f))  modes for $\Delta>0$ (the left column) and for
$\Delta<0$ (the right column) with the same parameters as in
Fig.~\ref{fig_w1w2}. In (a) and (b) the profiles of the first
component are shown for $\mu_1 = 0.86$ and $\mu_1 = 0.9$
(indicated by dashed lines in Fig.~\ref{fig_w1w2}),
correspondingly. In [(c),(d)] OS and in [(e),(f)] OA solutions for
the second component corresponding to branches C and B from
Fig.~\ref{fig_w1w2} are shown. Thick, thin, and dashed lines
correspond to $\mu_2 = 0.0; 0.05$ and $0.1$, respectively. The
dotted line in (a), (b) shows the periodic potential (thin and
dashed lines in (a) and (b) are not distinguishable on the scale of
the figure).} \label{fig_prof}
\end{figure}

We have introduced  Eqs.~(\ref{mode1}), (\ref{mode2_eigen}) for the
sake of approximation of localized modes in the limit $N_2\ll N_1$.
In order to construct the solutions emerging from the branches B and C we use an iterative procedure starting with the solutions obtained for $N_2\approx0$.  Fig.~\ref{fig_prof} shows examples of the
obtained results.

The mode for the second component bifurcates from the
zero-solution. Its amplitude grows with the detuning of the chemical potential
$\mu_2$ towards the center of the gap [see
Fig.~\ref{fig_prof}(c)-(f) and Fig.~\ref{fig_Veff} (a)],
 what is accompanied by the increase of the number of particles
$N_2$.
\begin{figure}[ht]
\epsfig{file=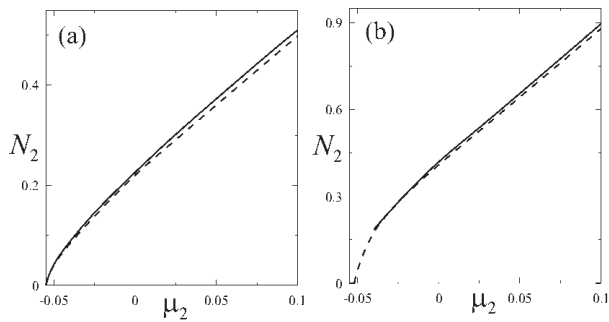,width=8cm}
\caption{
The number of particles
of the second component {\it vs} $\mu_2$ for the cases
$\Delta>0$ (panel a) and $\Delta<0$ (panel b) corresponding to
Fig.~\ref{fig_prof}. In (a) $\mu_1=0.86$ with $N_1\approx 3.17$ and
in (b) $\mu_1=0.9$ with $N_1\approx 3.42$. Here solid and dashed
lines correspond to second component with OS and OA symmetry.}
\label{fig_N_mu}
\end{figure}

In Fig.~\ref{fig_N_mu} we present the dependence of the number of
particles in the second component, $N_2$ on $\mu_2$ deviating from the
corresponding values B and C (see Fig.~\ref{fig_w1w2}). We notice
that in our procedure of search of the modes the profile of the
fundamental component also changes with $\mu_2$. The change, however, is
very small, it corresponds to the change of the number of particles in
the first component of order of  $10^{-3}$, and therefore is
practically invisible on the scale of Fig.~\ref{fig_prof}(a), (b).

To study dynamical stability of the obtained modes we performed
direct numerical integration of equations (\ref{GP_1D_1})  with
$m_1 = m_2$ and $\omega_1=\omega_2$. As initial conditions we used
perturbed stationary solutions obtained using the iterative
procedure, the amplitude of perturbation being of the order of
5$\%$.
\begin{figure}[h]
\epsfig{file=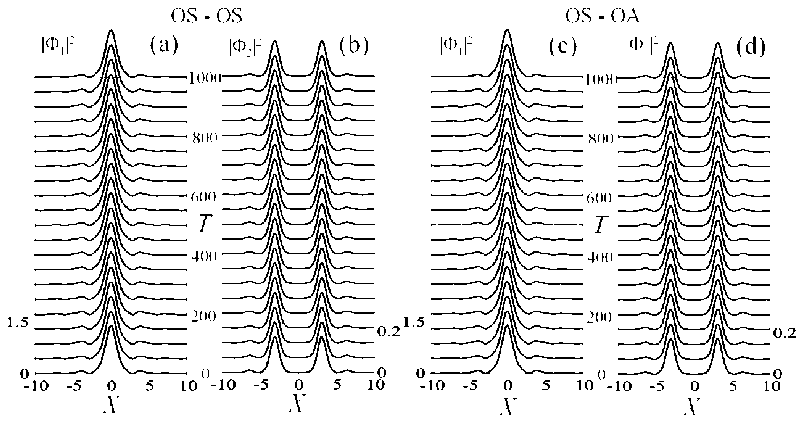,width=8cm} \caption{
Dynamics of density of the first component in (a), (c)  and of the
second component in (b), (d). In [(a), (b)] and [(c), (d)] initial
profiles are taken from Fig.~\ref{fig_prof} [(a), (c)] and [(a),
(e)], correspondingly, with  $\mu_1=0.86$ and $\mu_2=0.1$.}
\label{fig_dynamics_mag_par}
\end{figure}

\begin{figure}[h]
\epsfig{file=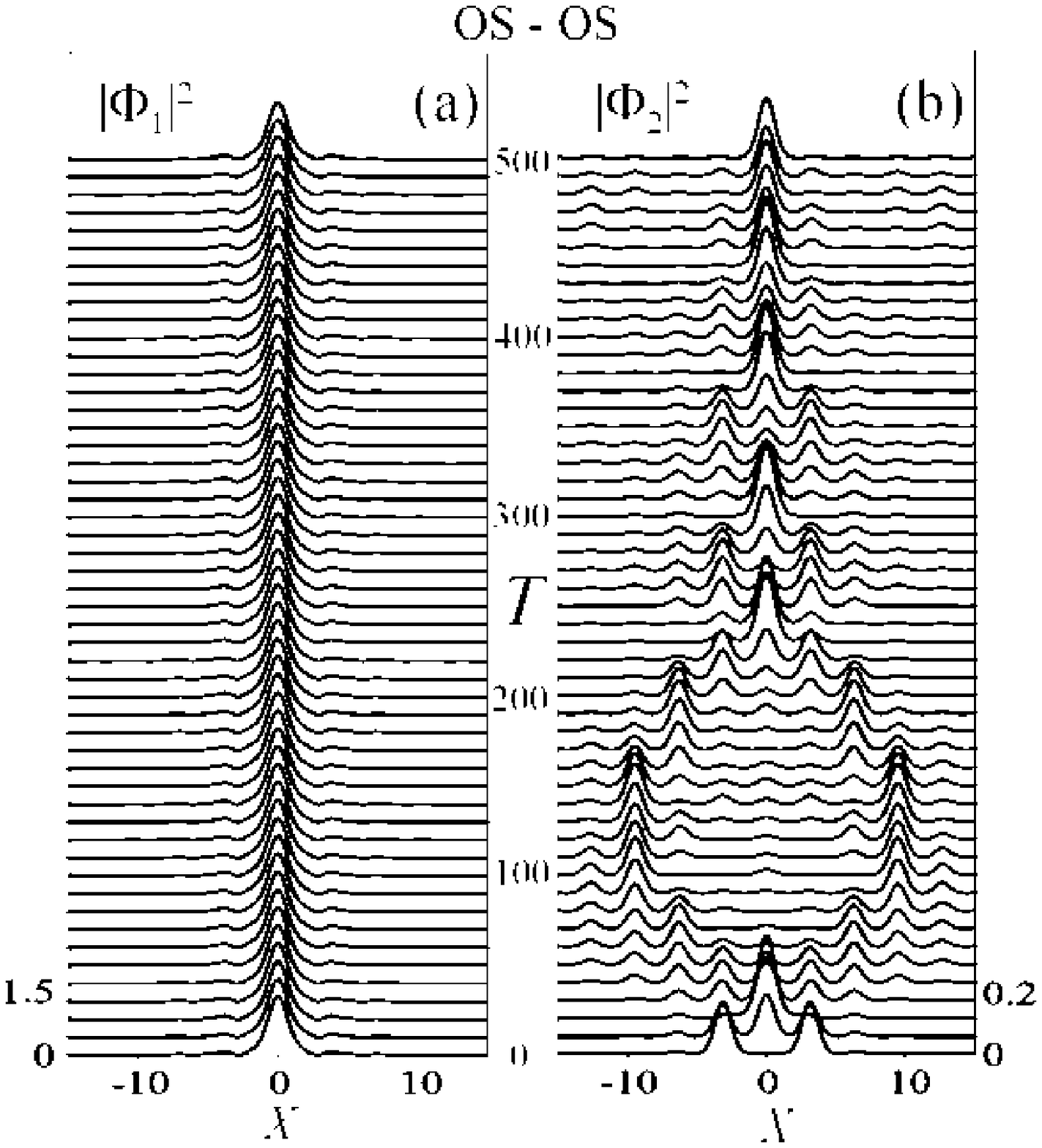,height=4cm,width=4cm}
\epsfig{file=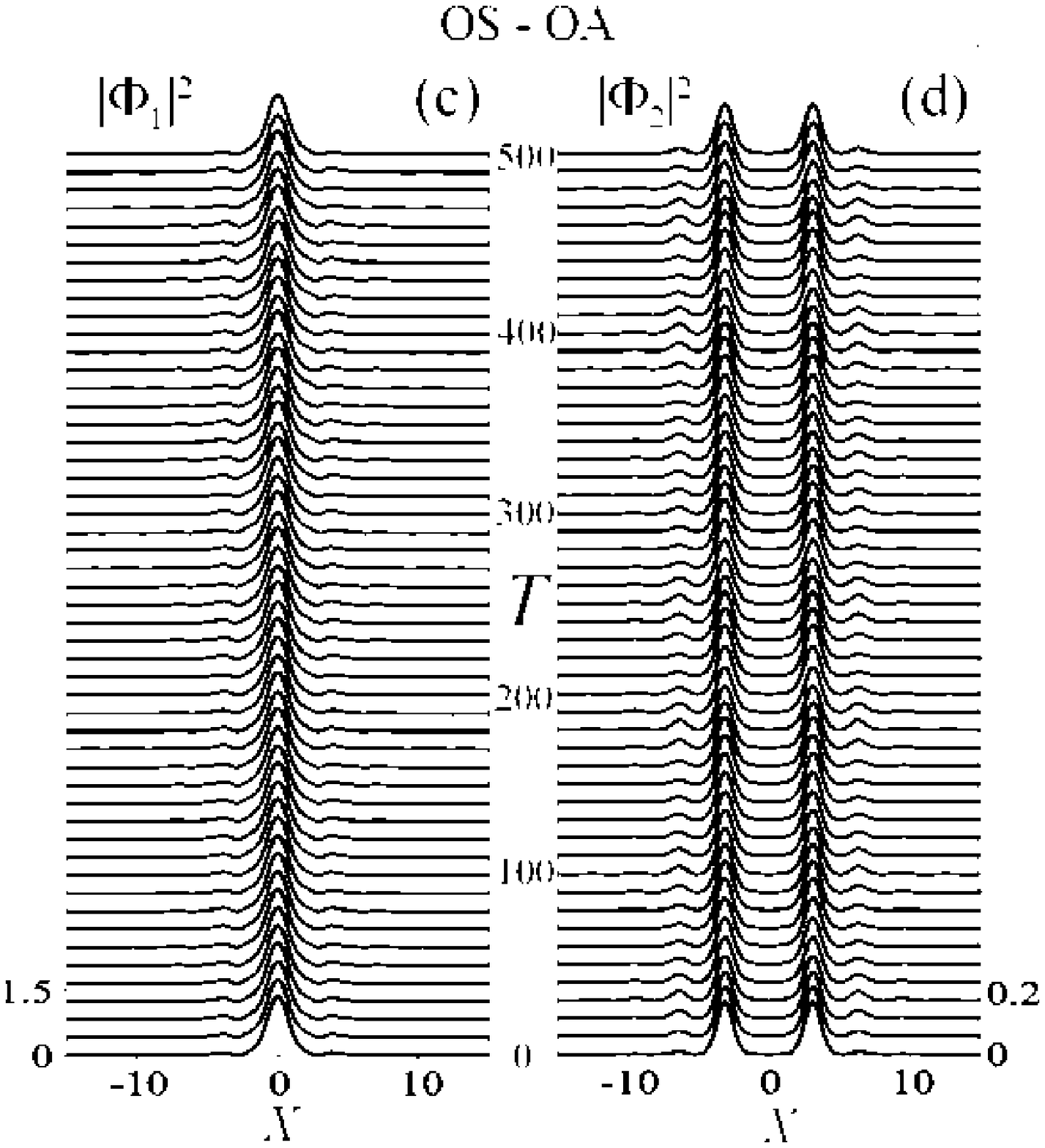,height=4cm,width=4cm} \caption{ The
same initial conditions as in Fig.~\ref{fig_dynamics_mag_par} with
$\chi=0$. The real domain of calculation is [-14$\pi$; 14$\pi$].}
\label{fig_dynamics_mag_chi0}
\end{figure}

First, in Fig.~\ref{fig_dynamics_mag_par}  we show the case
$\Delta>0$. One can see that the time evolution of the
two-component soliton is stable. By switching off the coupling
parameter, i.e. setting $\chi=0$, one obtains the system of two
uncoupled GP equations. The dominant fundamental component is
close, in some sense, to an OS mode of the single GP equation with
the periodic potential and that is why it is stable. The second
component,  representing two in-phase humps and originally coupled
by the effective potential due to the first component, becomes
unstable at $\chi=0$. The two humps in the absence of first
component, which the played role of the effective barrier between
them, start to move towards each other and overlap. However due to
the large velocity after collision two wave packets start to move outwards the
center. The described behavior is observed at the initial stages of 
the evolution shown in Fig.~\ref{fig_dynamics_mag_chi0} (a), (b). At
latter times, reaching the turning point they change the direction
of motion towards the center. After several periods of damped
oscillations the one-hump OS mode is created. 
The out-of-phase distribution of the second component, shown
in Fig.~\ref{fig_dynamics_mag_chi0} (d) is stable representing a
single-component stable OA mode. 



\begin{figure}[h]
\epsfig{file=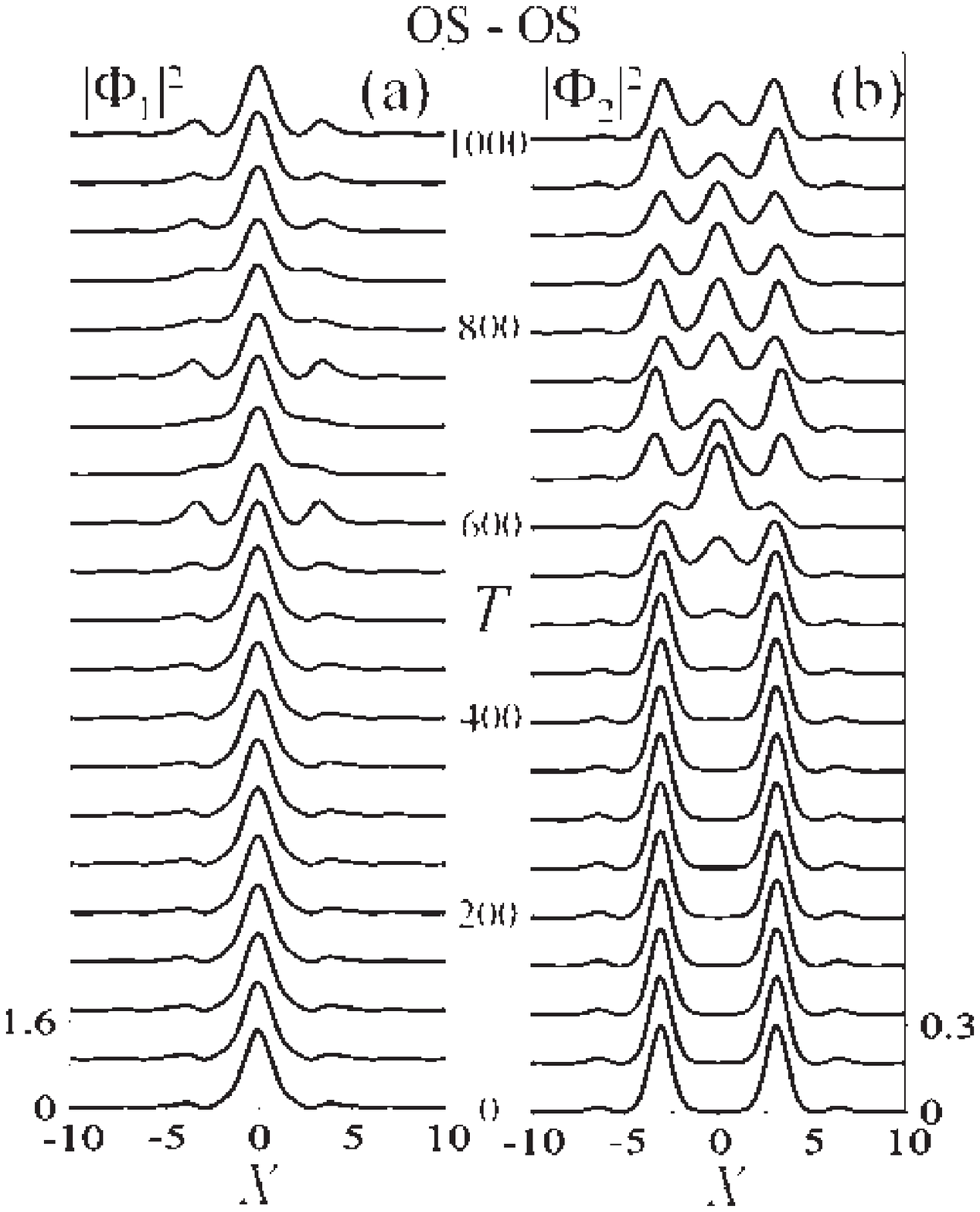,width=4cm}
\epsfig{file=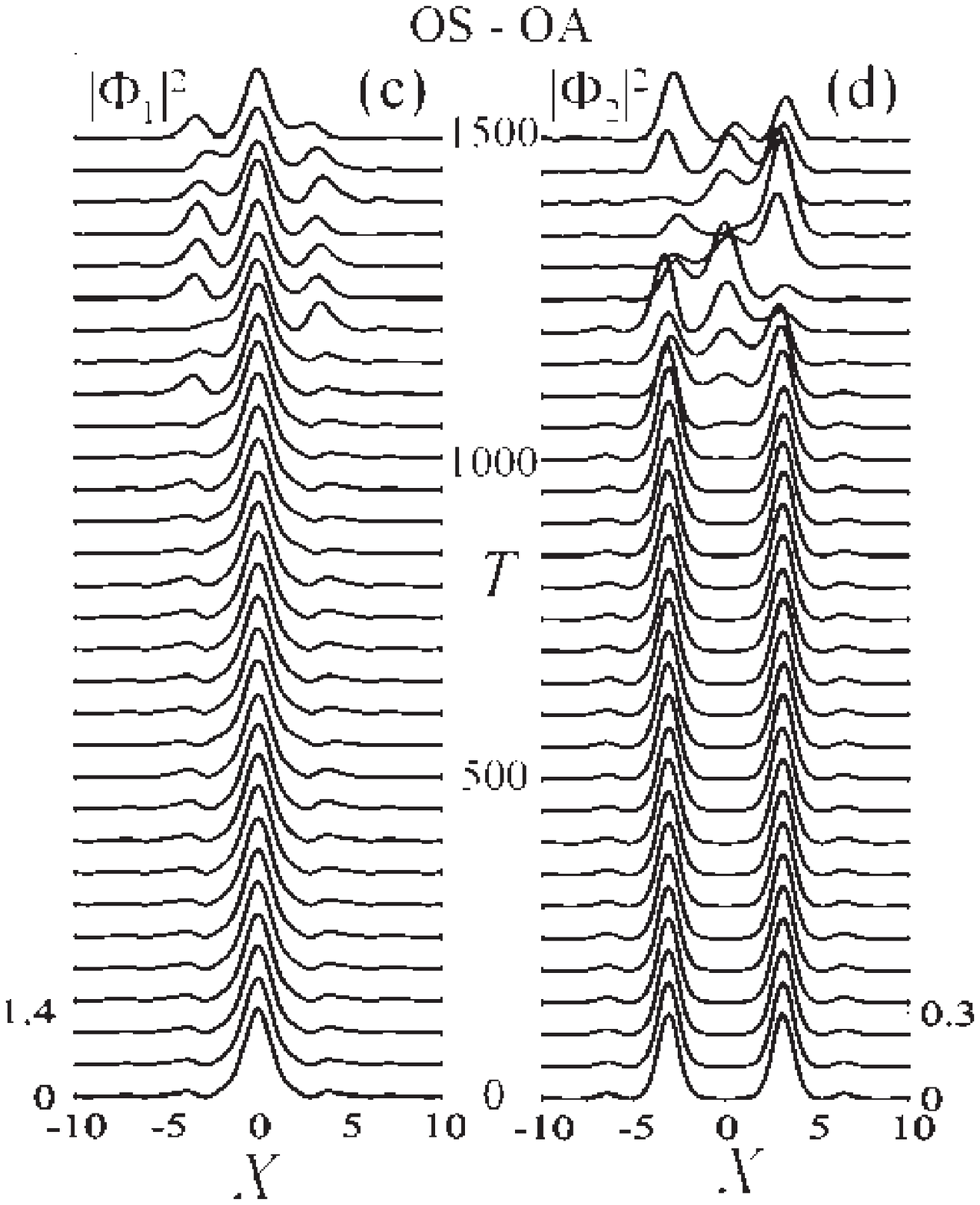,width=4cm}
\caption{ The dynamics of density profiles in (a), (c) of the first
component with OS symmetry and in (b), (d) of the second component with OS and OA
symmetries. Initial distributions are as in  Fig.~\ref{fig_prof} [(b),(d)] and [(b),(f)] with $\mu_1=0.9$ and
$\mu_2=0.1$, respectively.}
\label{fig_dynamics_opt_par}
\end{figure}

Next we investigated the case $\Delta<0$. Typical results are
shown in Fig.~\ref{fig_dynamics_opt_par}, where one observes
instability of the both OS-OS (panels (a) and (b)) and OS-OA (panels
(c) and (d)) modes. It has been checked that the time at which the
instability is developed increases when $\mu_2$ approaches the
lowest edge of the gap, tending to infinity as $\mu_2$ approaches
$\mu_{B,C}$ and respectively as $N_2\to 0$.


\section{Localized modes with balanced number of atoms (limit $N_1\simeq N_2$).}
\label{LargeAmpl}

Let us now turn to the study of two-component localized modes with
{\it a balanced}  i.e. with the same or with similar number of
particles in each component. These modes can be of  equal symmetry
or of mixed symmetry type. Some of the equal symmetry modes were
also found in Ref. \cite{2comp_localization}. Balanced mixed symmetry modes
have not been discussed before. In contrast to the equal symmetry
solutions, which for many aspects are similar to standard modes of
the single component GPE, the mixed state symmetries are more
specific for multicomponent BEC systems in OL displaying
interesting stability properties.
\begin{figure}
\centerline{
\includegraphics[width=4.25cm,height=4.25cm,clip]{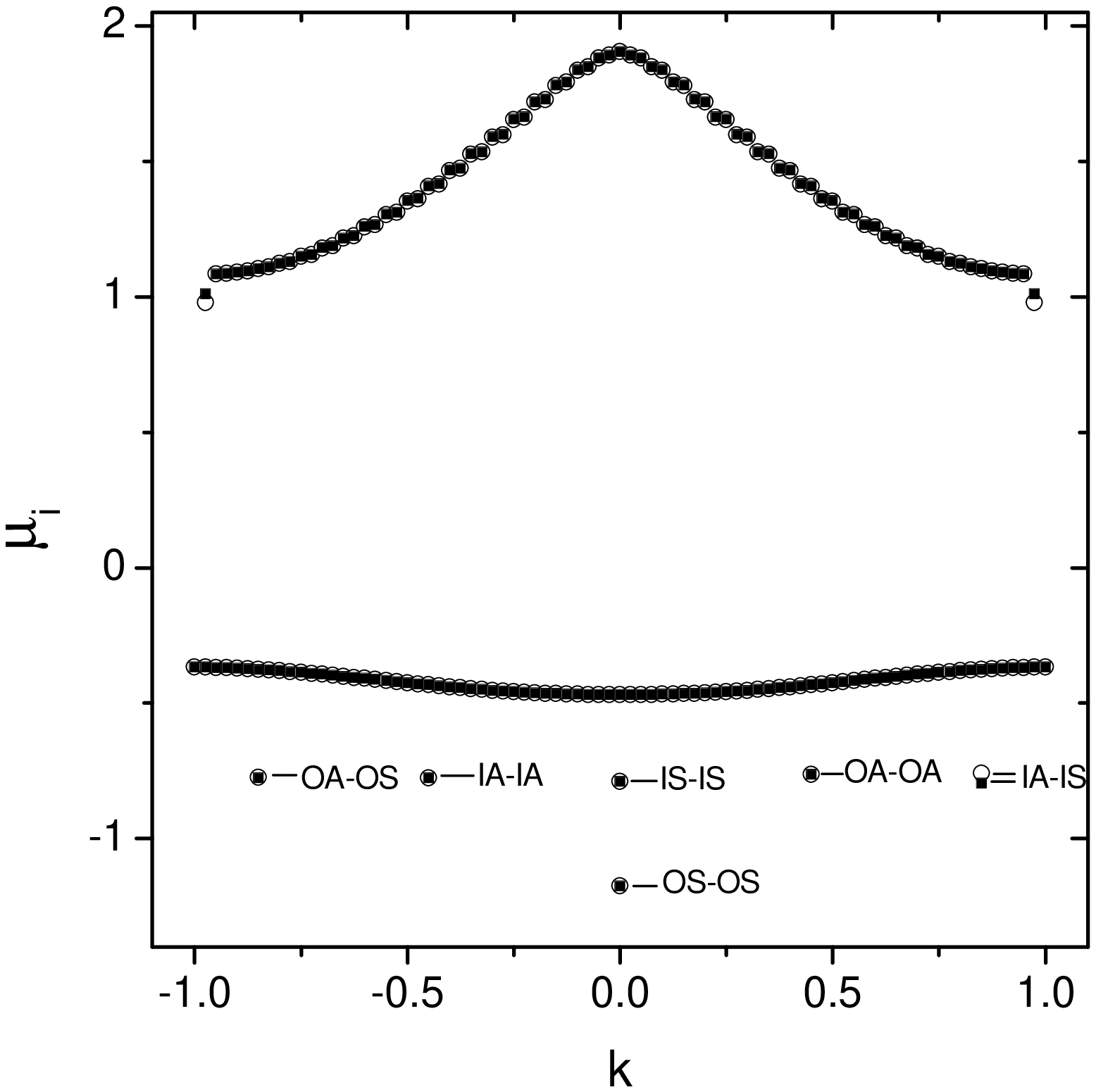}
\includegraphics[width=4.25cm,height=4.25cm,clip]{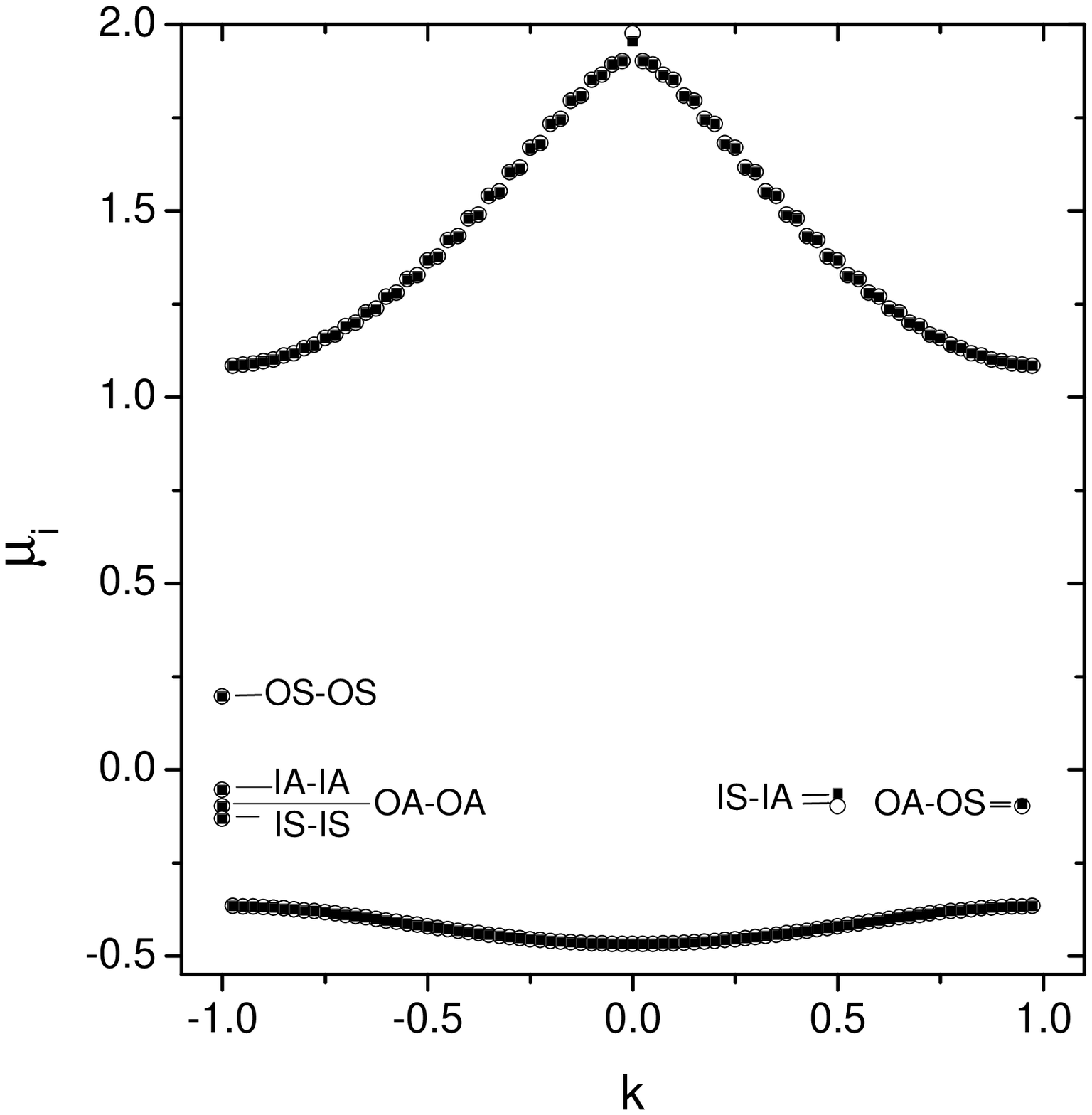}
} \caption{Band structure  and bound state levels of the two
component BEC in OL  with all attractive
$\chi_{11}=\chi_{22}=\chi_{12}=-0.5,$ (left panel) and all
repulsive $\chi_{11}=\chi_{22}=\chi_{12}=0.5,$ (right panel)
interactions. For both cases the amplitude of the OL
is $V_0=1.5$ and the number of atoms is $N_1=1.5, N_2=0.98$. Open
circles refer to the first component while filled squares refer to
the second component. Notice that all equal symmetry modes are
degenerated. Also  notice the presence of a bound state near the
edge of the upper band (left panel )and a bound state n the second
gap (right panel). \label{fig1ms}}
\end{figure}
\begin{figure}
\centerline{
\includegraphics[width=2.65cm,height=2.65cm,angle=0,clip]{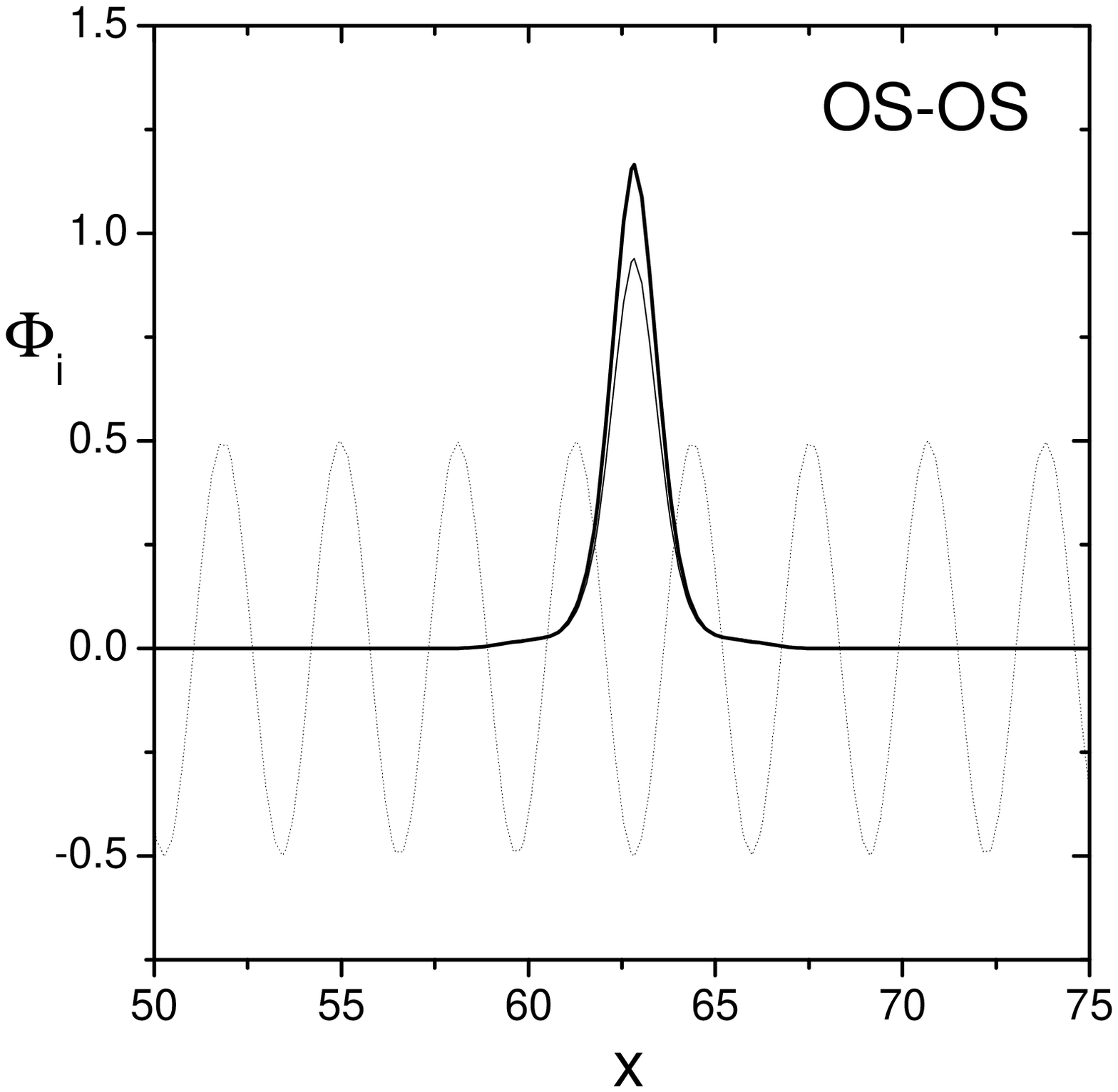}
\includegraphics[width=2.65cm,height=2.65cm,angle=0,clip]{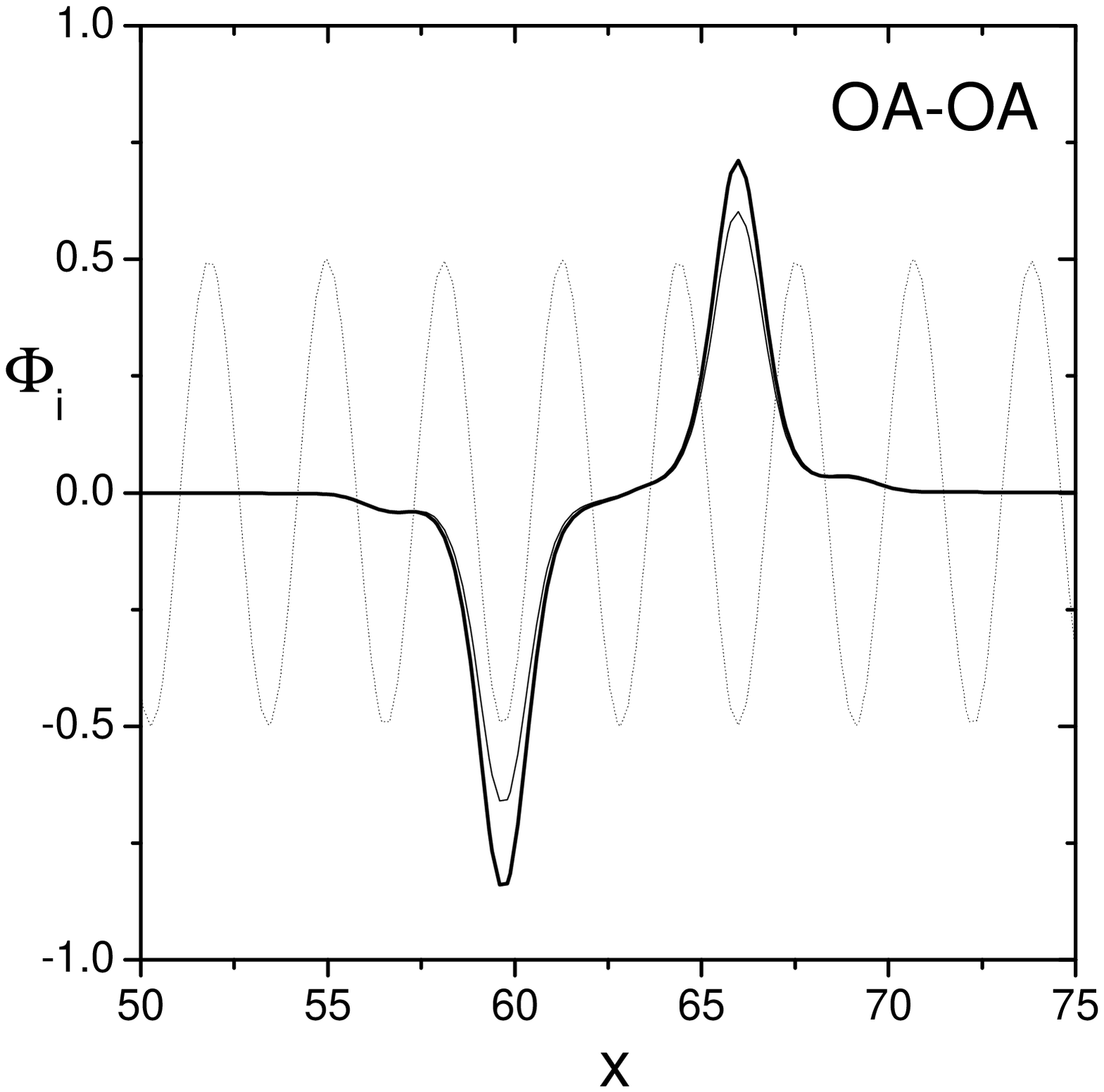}
\includegraphics[width=2.65cm,height=2.65cm,angle=0,clip]{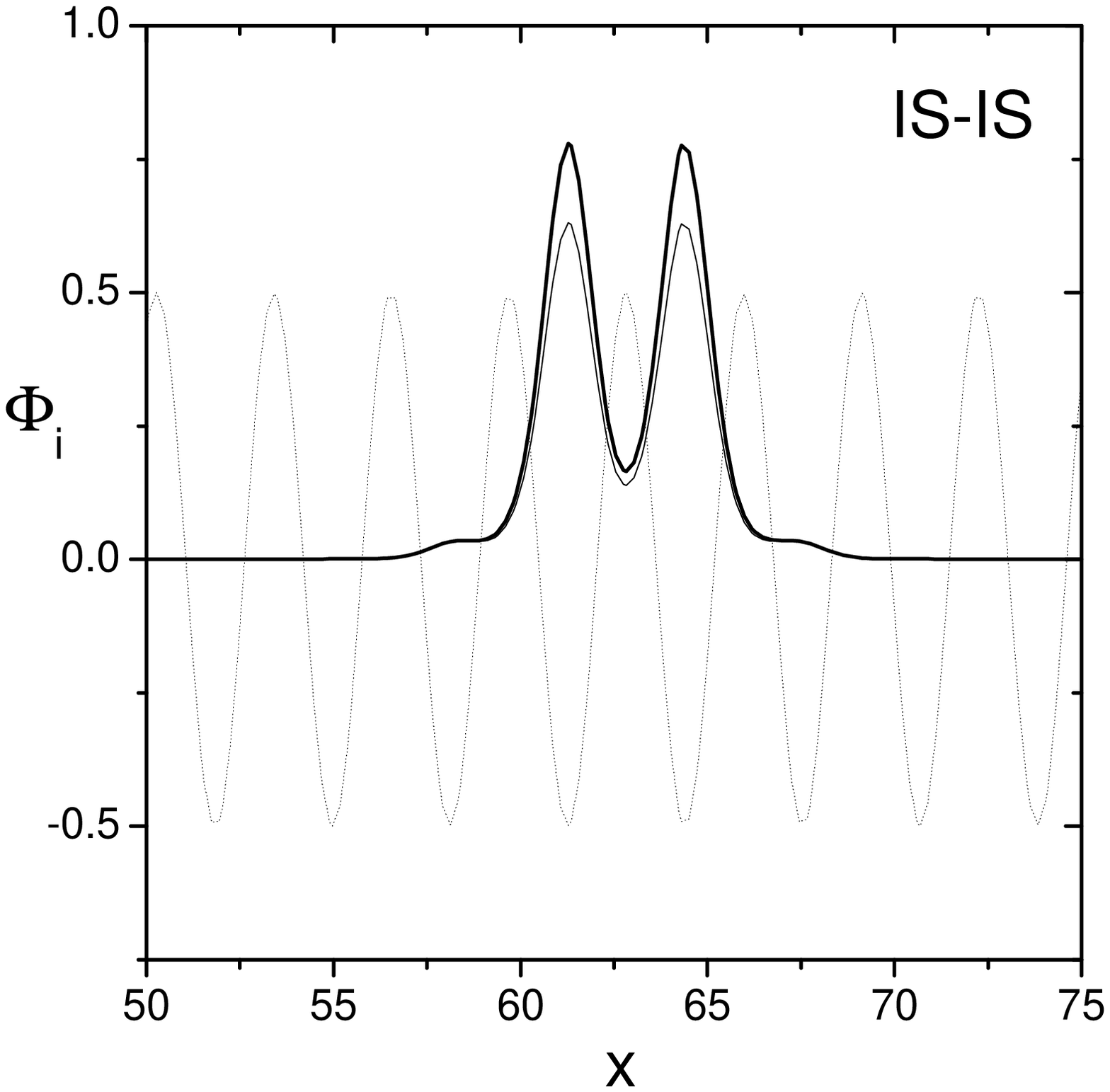}}
\centerline{
\includegraphics[width=2.65cm,height=2.65cm,angle=0,clip]{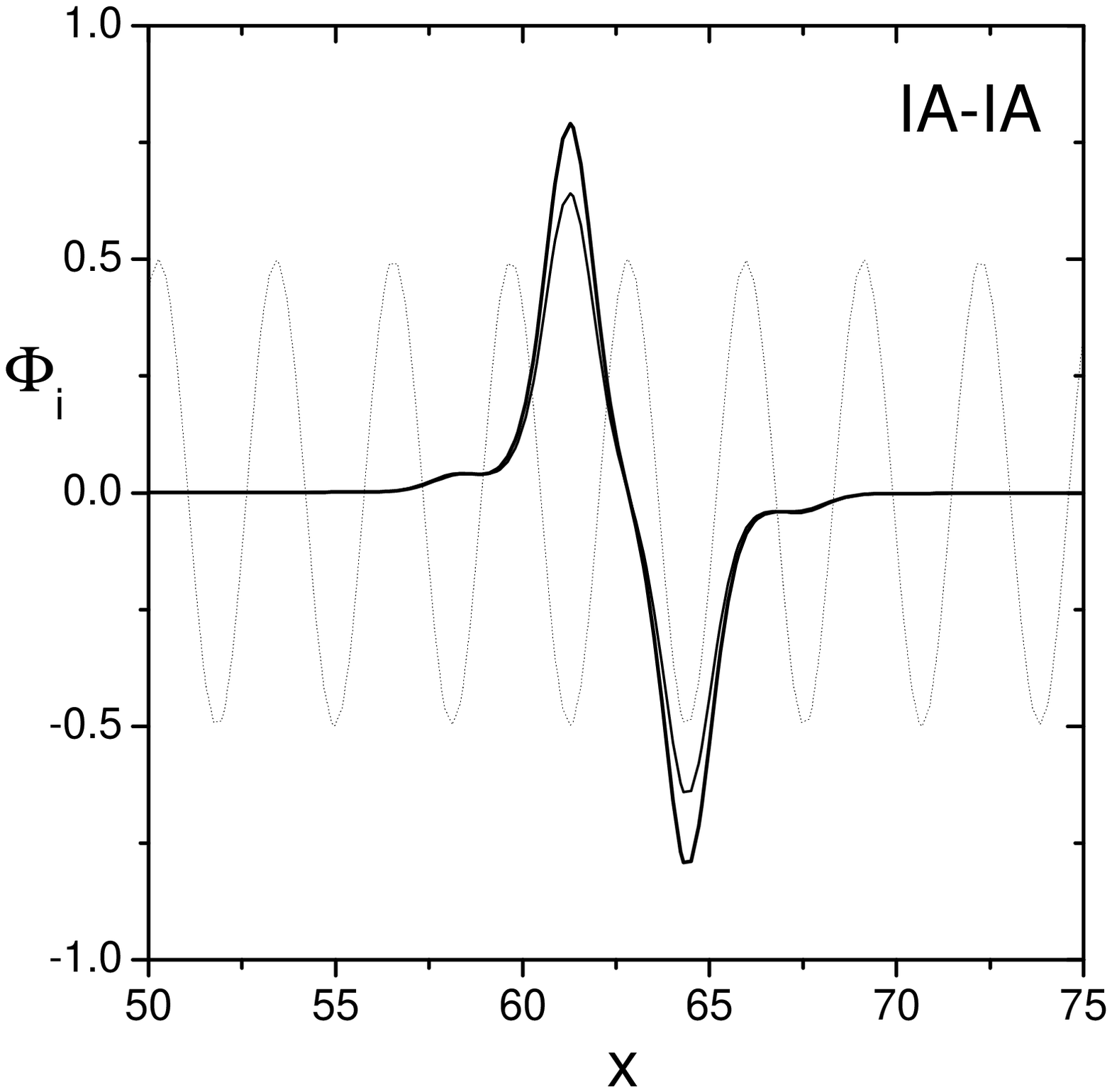}
\includegraphics[width=2.65cm,height=2.65cm,angle=0,clip]{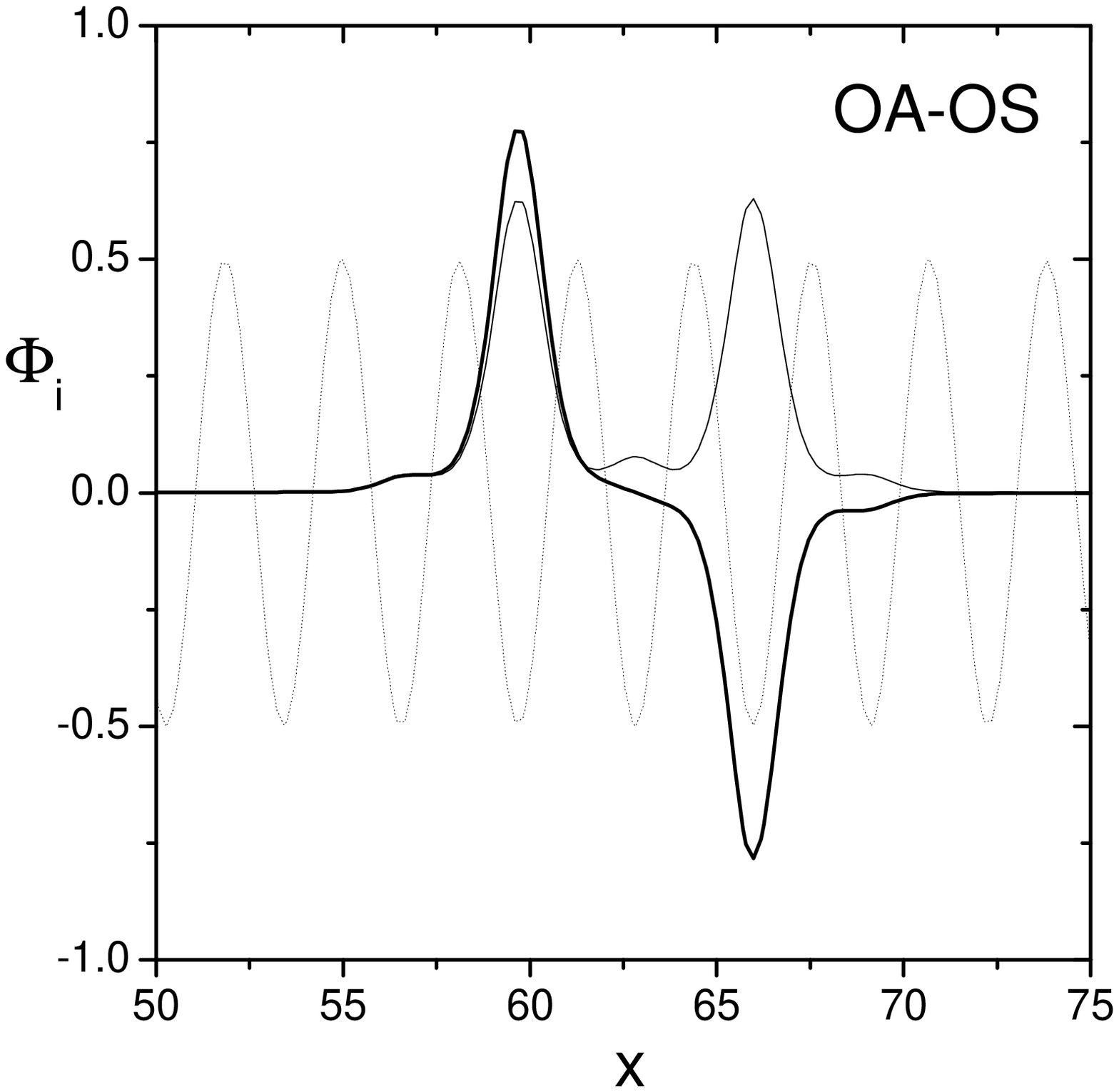}
\includegraphics[width=2.65cm,height=2.65cm,angle=0,clip]{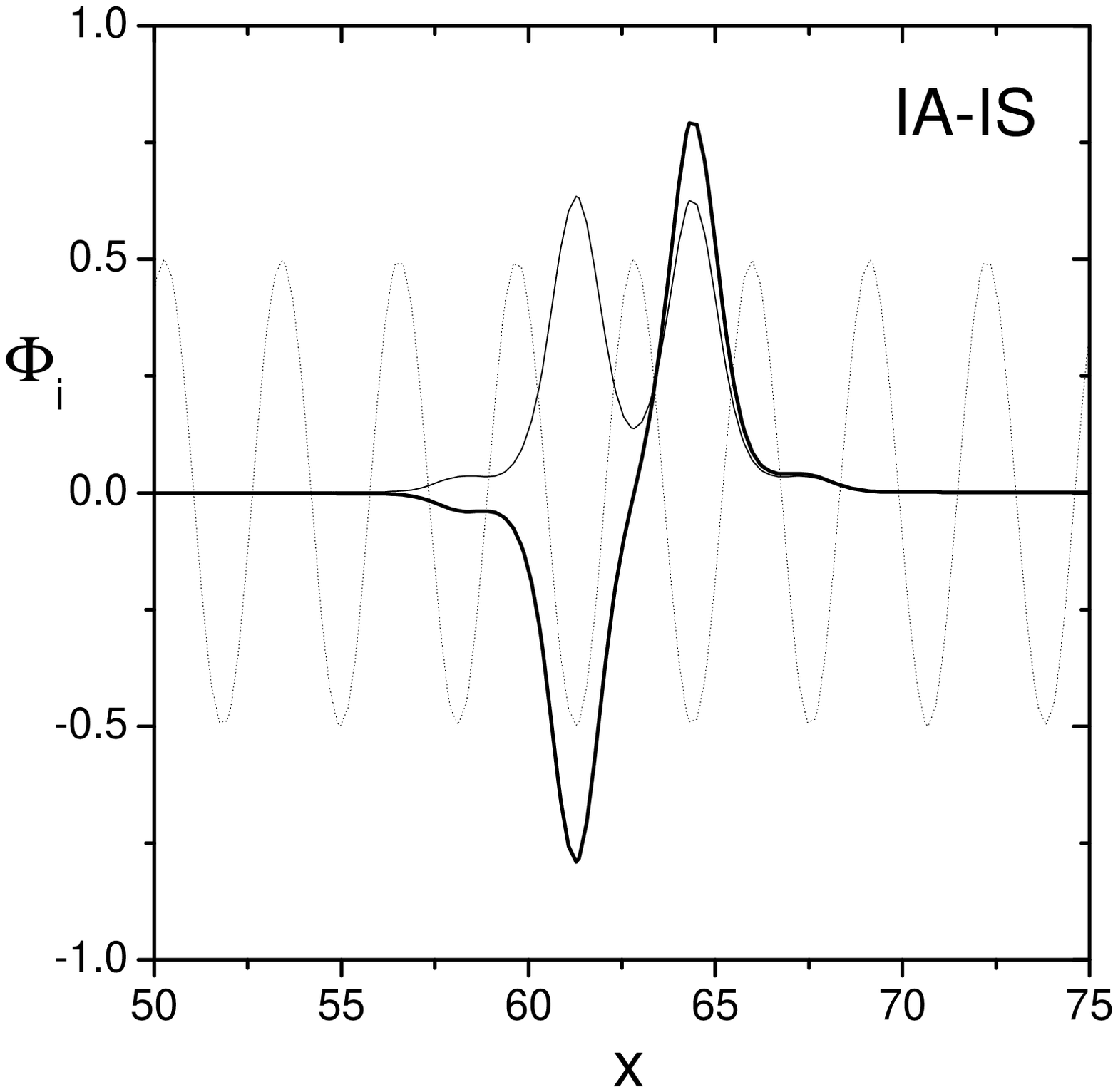}}
\caption{Condensate wavefunctions of two-component BEC in OL
corresponding to the bound state levels in the left panel of Fig.
\ref{fig1ms} (all attractive case). Thick and thin lines refer to
first and second component, respectively (open circles and filled
squares in Fig.~\ref{fig1ms}). To identify the symmetry of the
solutions, lattice potentials (scaled by a factor 3) have been
shown by thin dotted line.} \label{fig2ms}
\end{figure}
\begin{figure}
\centerline{
\includegraphics[width=2.65cm,height=2.65cm,angle=0,clip]{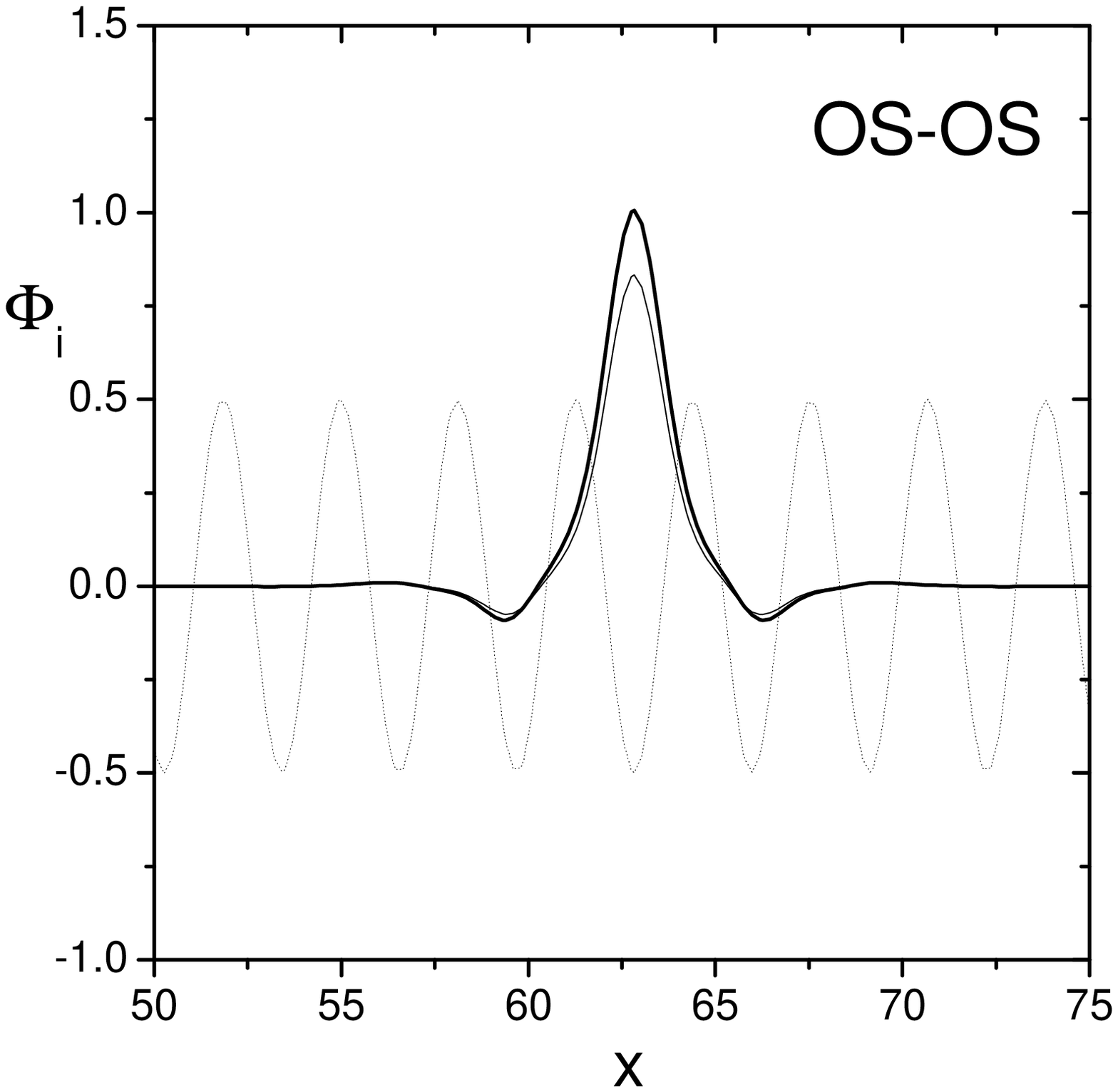}
\includegraphics[width=2.65cm,height=2.65cm,angle=0,clip]{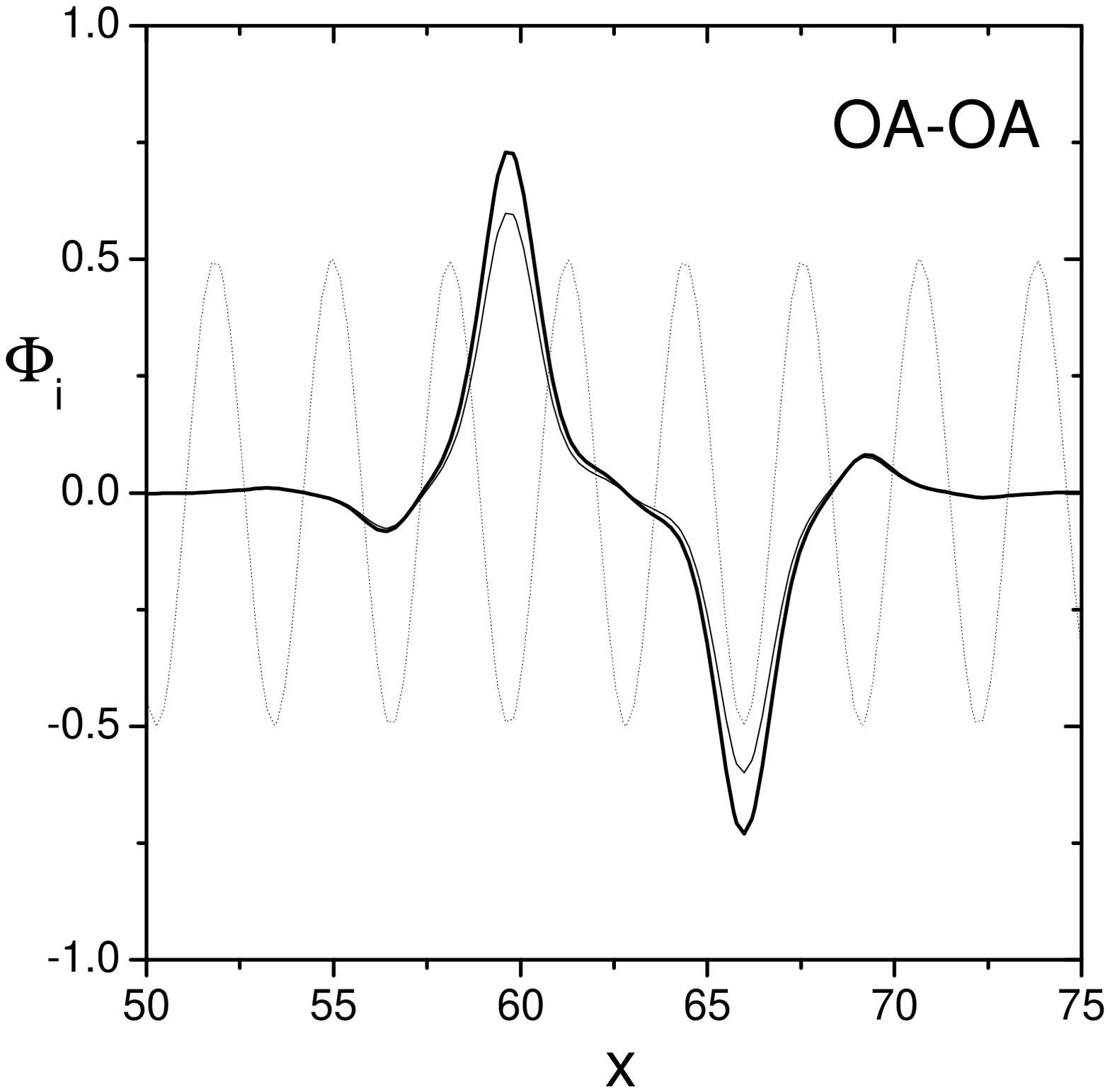}
\includegraphics[width=2.65cm,height=2.65cm,angle=0,clip]{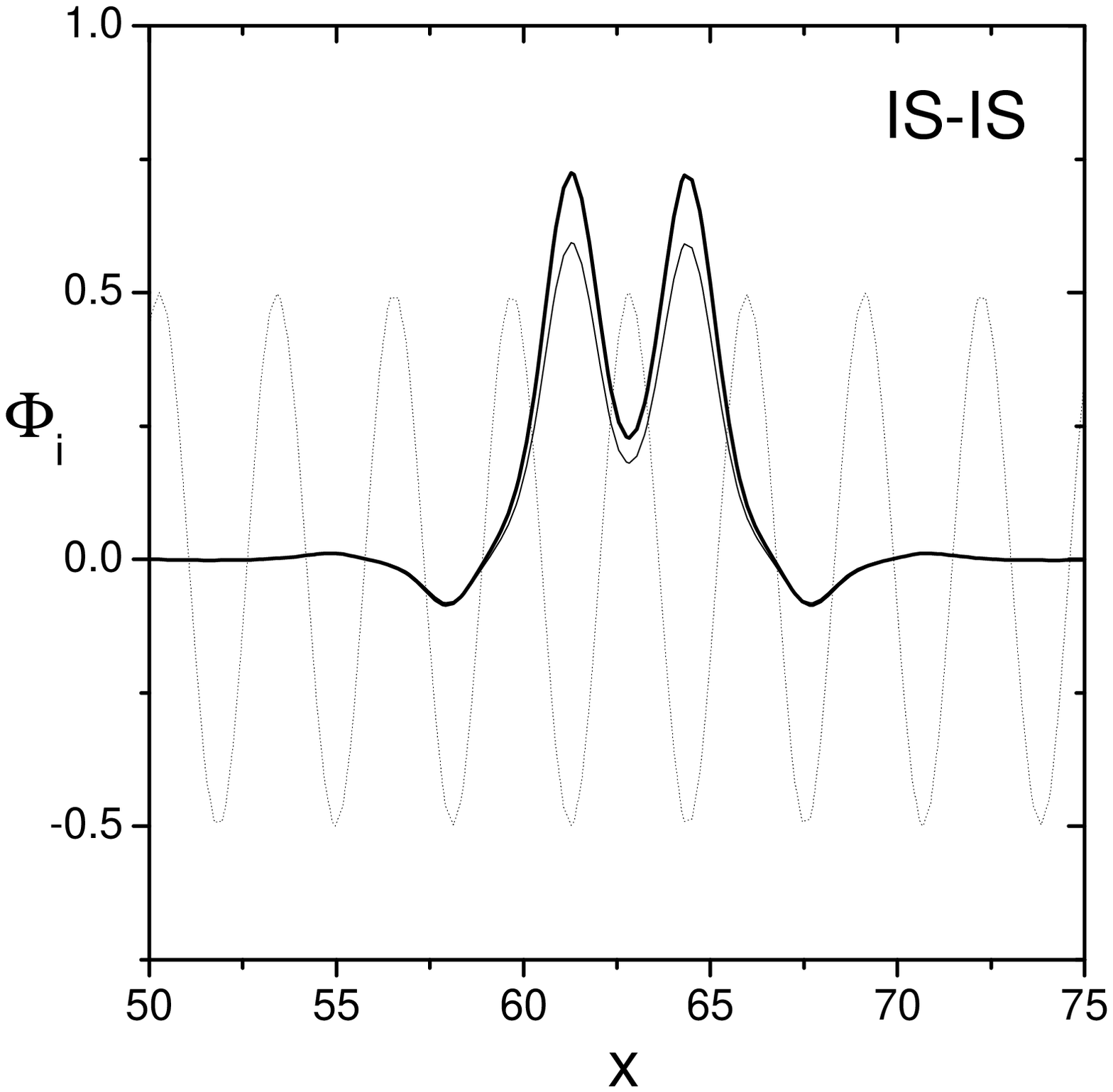}}
\centerline{
\includegraphics[width=2.65cm,height=2.65cm,angle=0,clip]{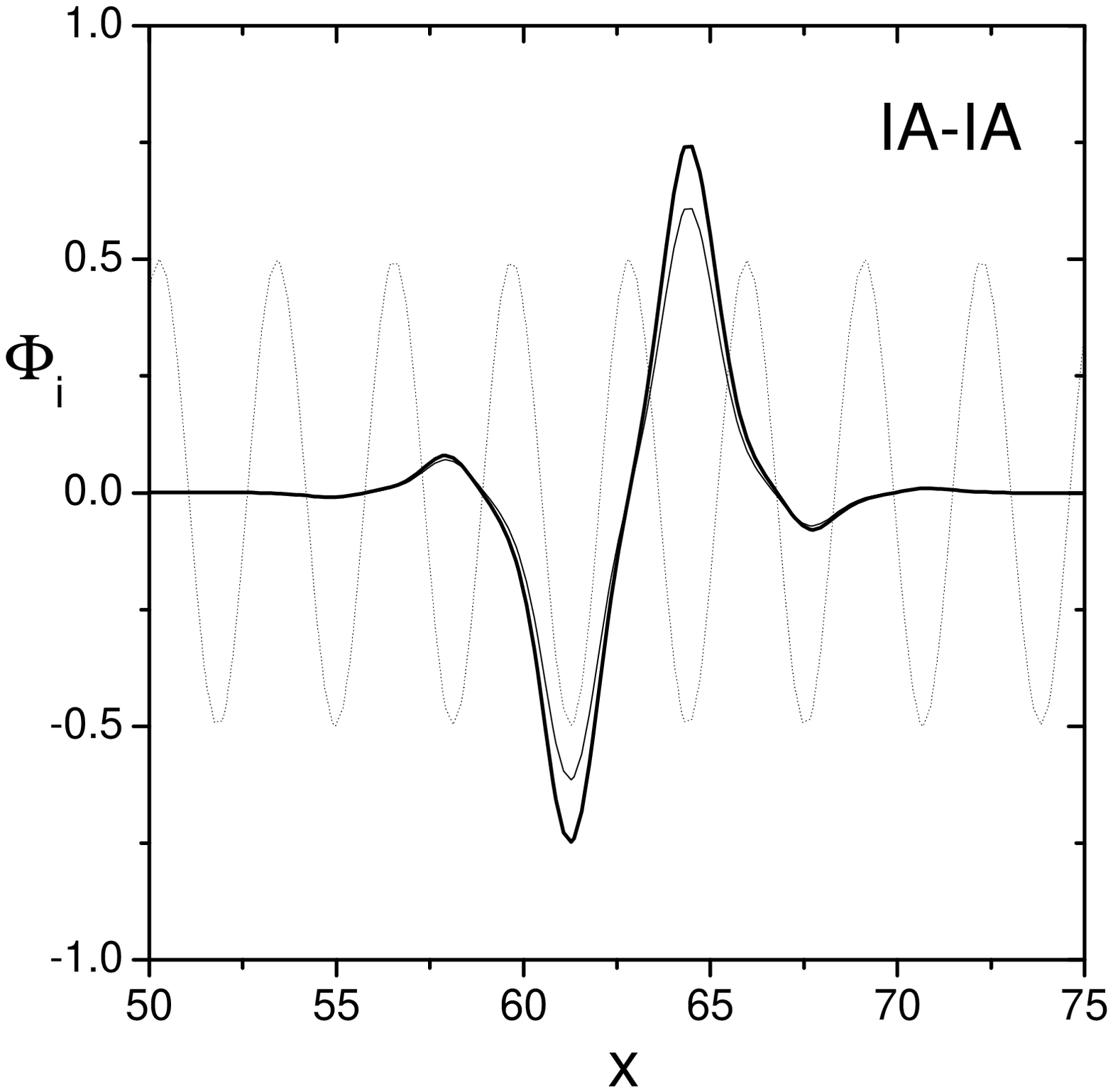}
\includegraphics[width=2.65cm,height=2.65cm,angle=0,clip]{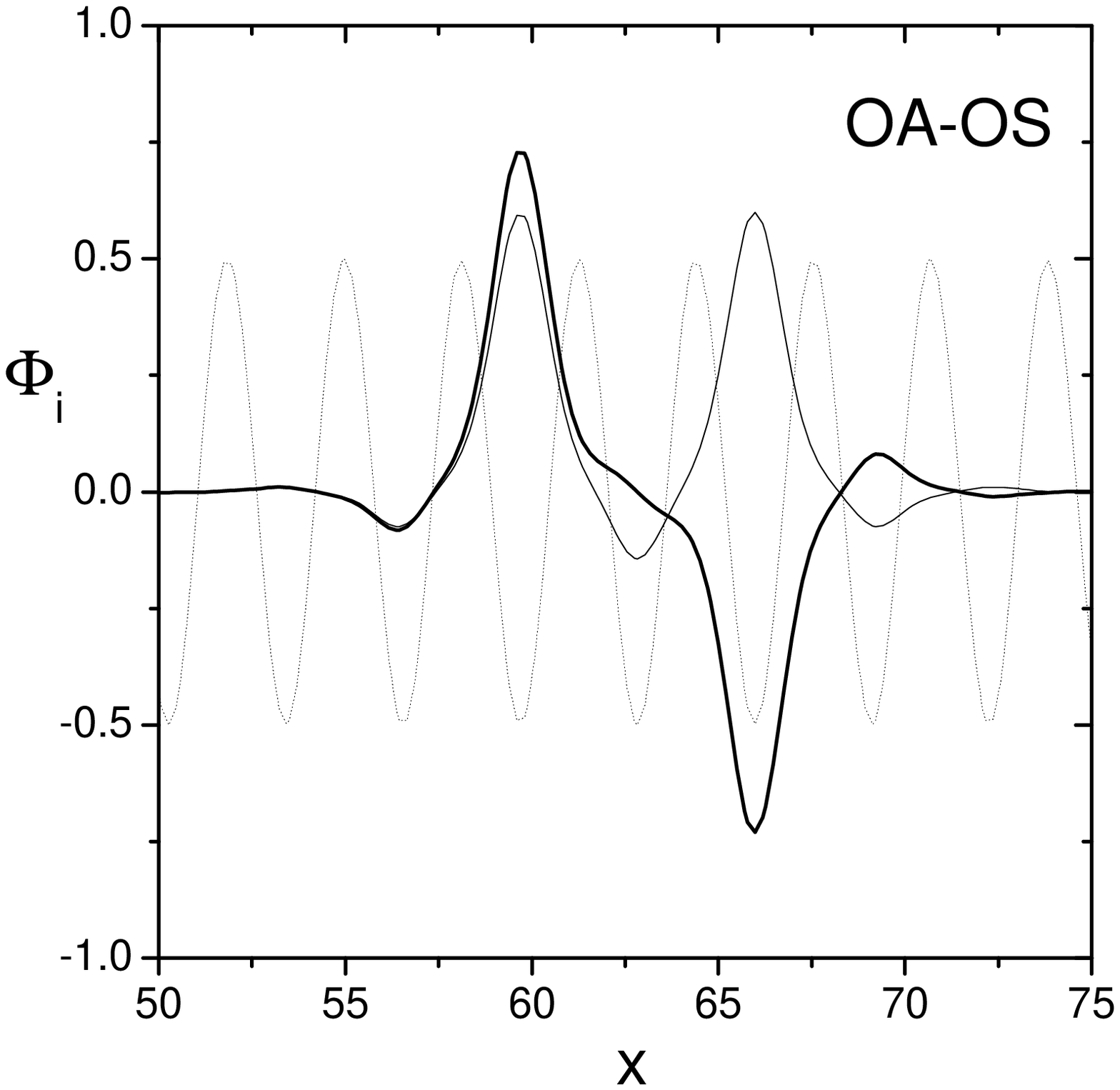}
\includegraphics[width=2.65cm,height=2.65cm,angle=0,clip]{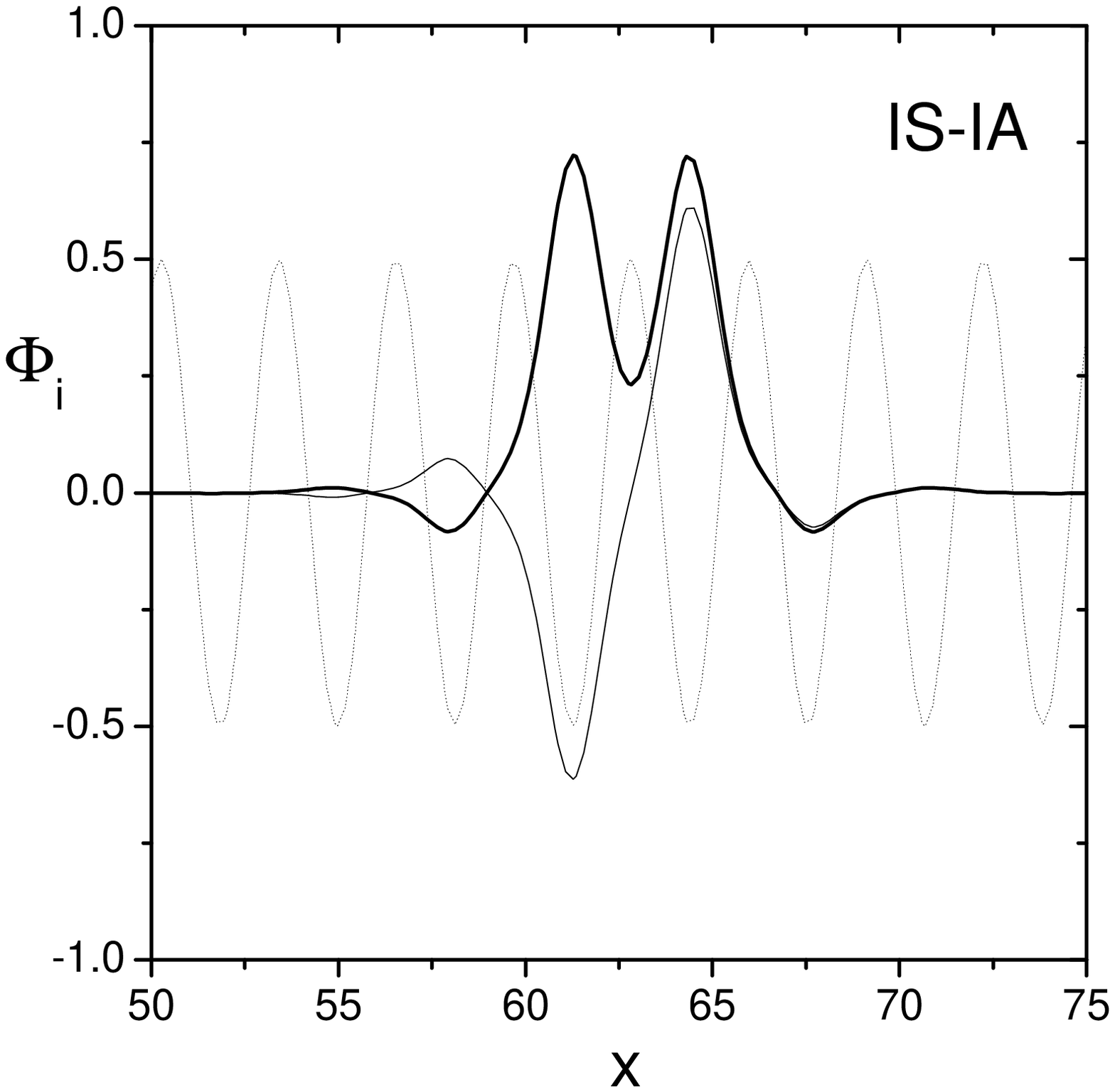}}
\caption{Condensate wavefunctions of two-component BEC in OL
corresponding to the bound state levels in the right panel of Fig.
\ref{fig1ms} (all repulsive case). Thick and thin lines refer to
first and second component, respectively (open circles and filled
squares in Fig.~\ref{fig1ms}). To identify the symmetry of the
solutions, lattice potentials (scaled by a factor 3) have been
shown by thin dotted lines.}\label{fig3ms}
\end{figure}
Thus,  while the OA and the OA-OA modes are unstable for
the one- and two-component system, respectively,  the coupling of
the OA mode with an OS mode gives rise to a stable OS-OA mixed
state of two-component BEC in OL. The same  situation occurs with
the IS and IA modes. In the following we show that these mixed
symmetry solutions exist for both attractive and repulsive inter
and intra specie interactions and together with the on site OS-OS
solution, they represent stable localized excitation of a two-component BEC in OL.
\begin{figure}[htb]
\centerline{
\includegraphics[width=4.2cm,height=5.5cm,clip]{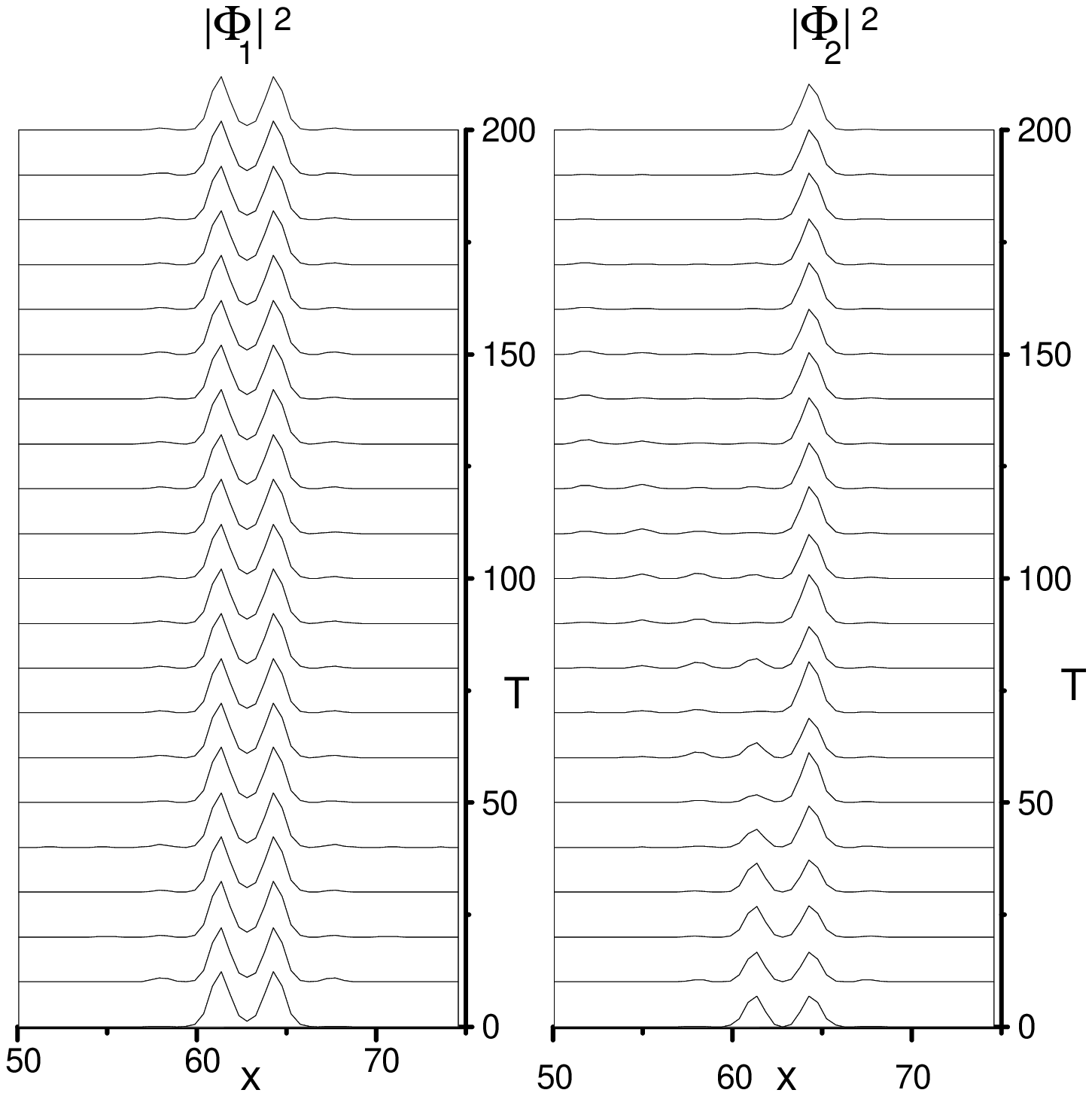} \
\includegraphics[width=4.2cm,height=5.5cm,clip]{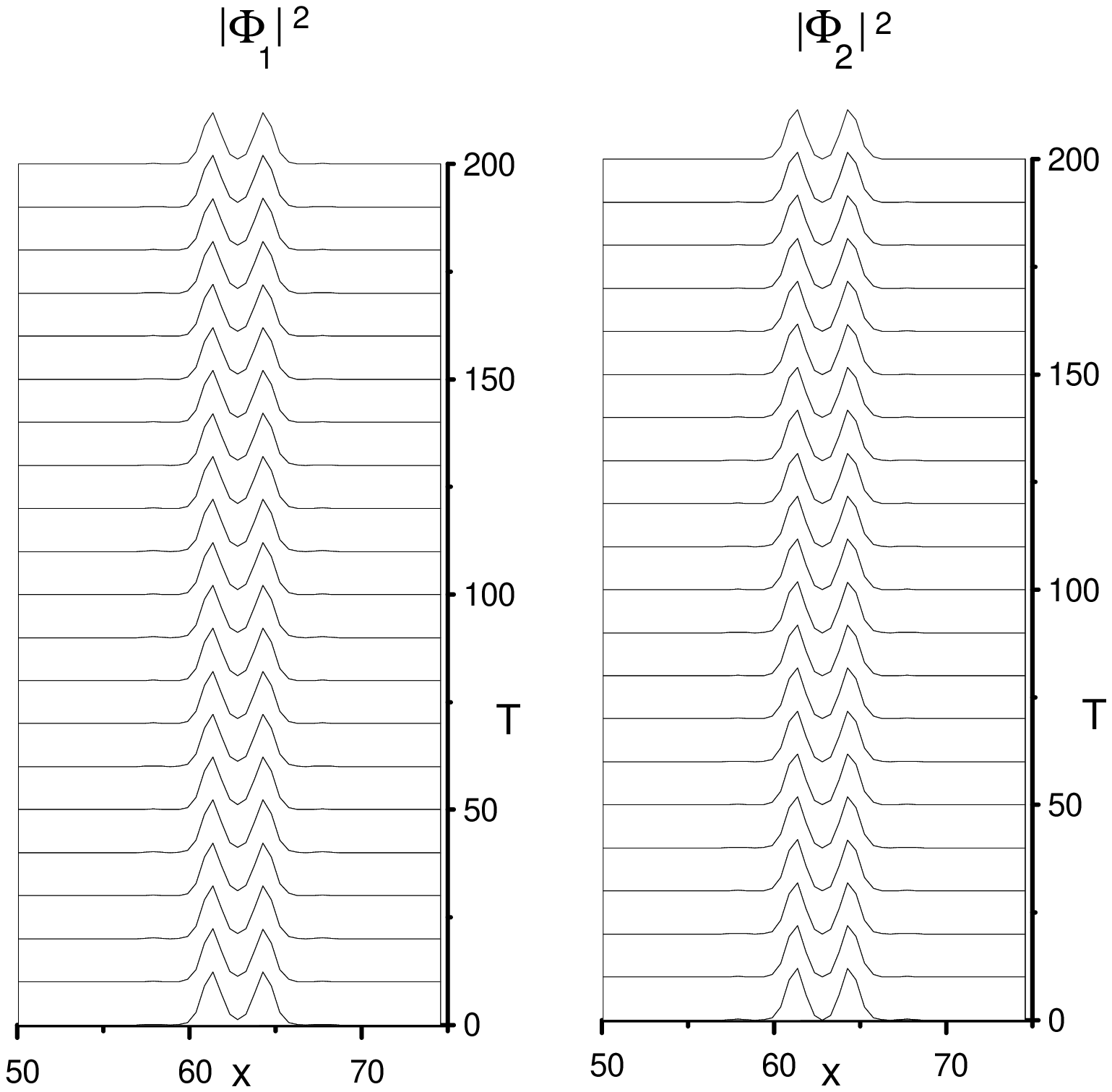}}
\caption{Time evolution of the density profiles of the IS-IA mode
of the all attractive case in Fig.~\ref{fig2ms}, in absence
($\chi=0$, left panels) and in presence ($\chi=-0.5$, right
panels) of interaction between the two components.} \label{fig4ms}
\end{figure}
\begin{figure}[htb]
\centerline{
\includegraphics[width=4.25cm,height=5.5cm,clip]{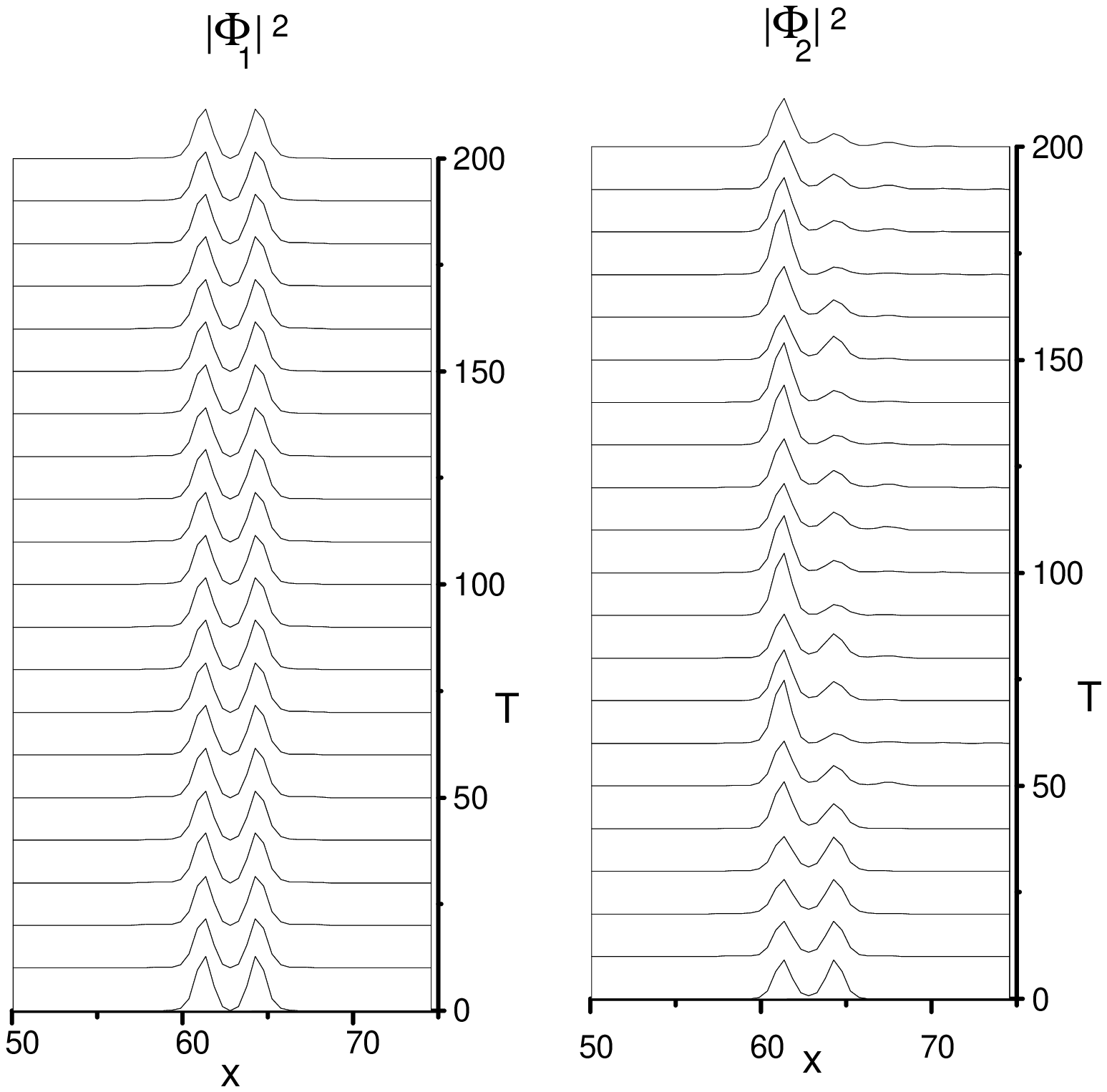} \
\includegraphics[width=4.25cm,height=5.5cm,clip]{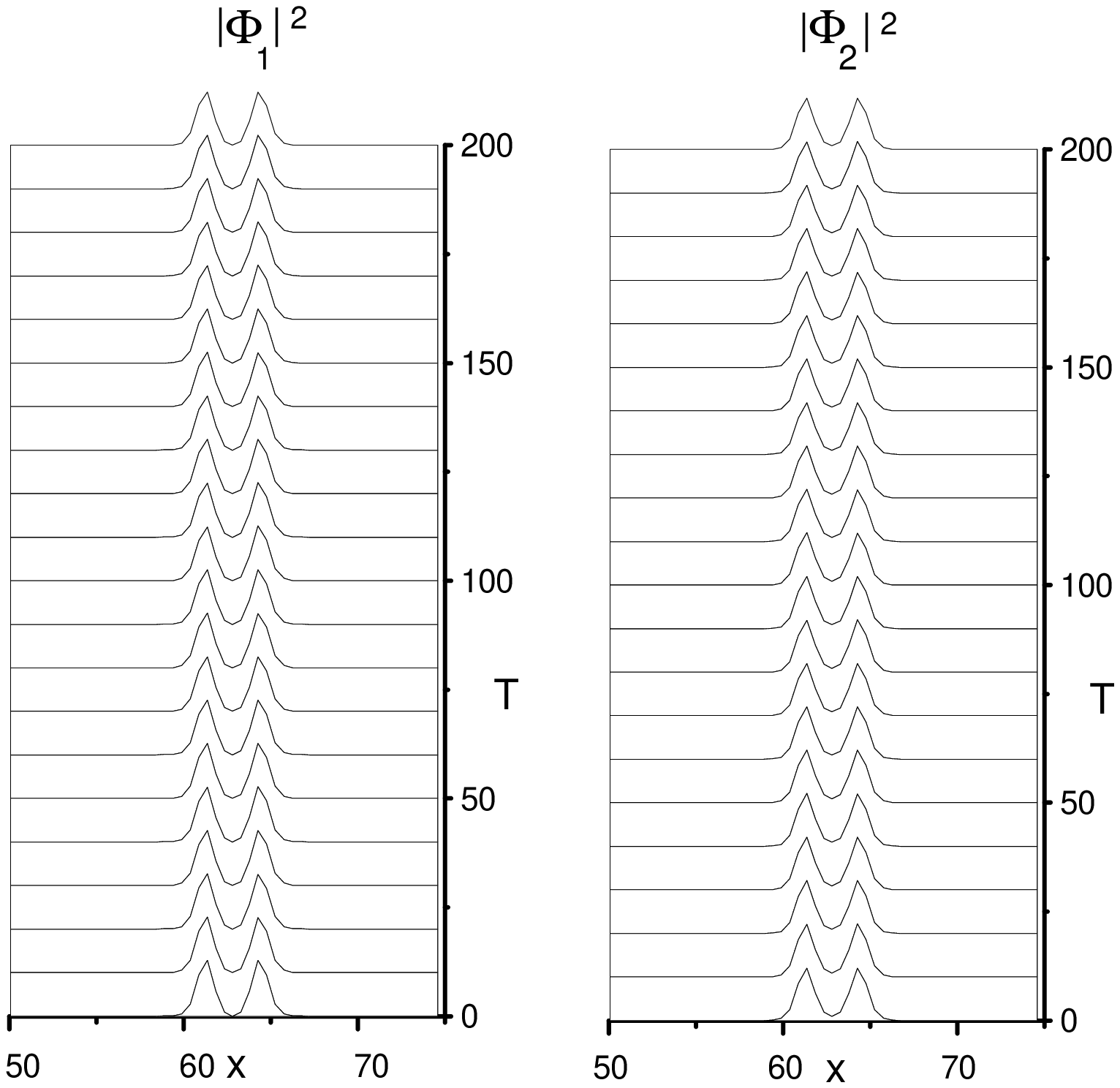}}
\caption{The same as in Fig.~\ref{fig4ms} but for the IS-IA mode of
the all repulsive case in Fig.~\ref{fig3ms}.}  \label{fig5ms}
\end{figure}

In Fig.~\ref{fig1ms} we depict the lowest part of the band
structure of Eqs.~(\ref{eigen_spinor}) obtained for the cases of
all attractive $\chi_{11}<0,\chi_{22}<0,\chi<0$ (left panel) and
all repulsive $\chi_{11}>0,\chi_{22}>0,\chi>0$ (right panel)
interactions. We see that the nonlinearity induces in both cases a
number of bound states in the band gap structure, located in the
forbidden zone  below the first band for the attractive case or in
the gap between the first two bands for the repulsive case.
Similar bound states can also form inside the upper band gaps (see
right panel of Fig.~\ref{fig1ms}). In this figure bound state
levels are labeled by symmetry type symbols close to the pair of
the corresponding levels. Thus the OS-OS symbol denotes a bound
state consisting of an OS state in both components with the
corresponding chemical potentials labeled by the open circle and
the full square close to the symbol. We find that the two
component bound states of same symmetry type are degenerate in
energy (chemical potential) both for all attractive and all
repulsive interactions.  In contrast to the states OS-OS and OA-OA
which are found to be stable also in absence of the interspecies
interaction ($\chi=0$), the states IA-IA is stable only in
presence of interaction between the two species and the state
IS-IS is unstable also in presence of the interspecies interaction
(in this case the state decays a bit faster without interaction).
In Figs.~\ref{fig2ms} and \ref{fig3ms} we depict the eigenstates
corresponding to the bound states shown in Fig.~\ref{fig1ms}.

In addition to the equal symmetry two-component bound states there
are also states with mixed symmetry such as the OA-OS and the
IS-IA bound states. These states are very stable and represent new
excitations of the two-component system since they can exist
only in presence of the interspecie interaction. In particular we
find that the mixed symmetry states OA-OS and IA-IS exist and are
stable under time evolution  only due to the interspecie
interaction. This is shown in Figs.~\ref{fig4ms} and \ref{fig5ms}
for the IS-IA mode of the attractive and repulsive cases in Fig.
\ref{fig2ms} and Fig.~\ref{fig3ms}, respectively. We see that in
both cases modes are stable in both components only in presence of
interaction and the switching off of the interaction makes the
second component to decay into an OS mode with emission of matter.
\begin{figure}
\centerline{
\includegraphics[width=4.25cm,height=4.25cm,clip]{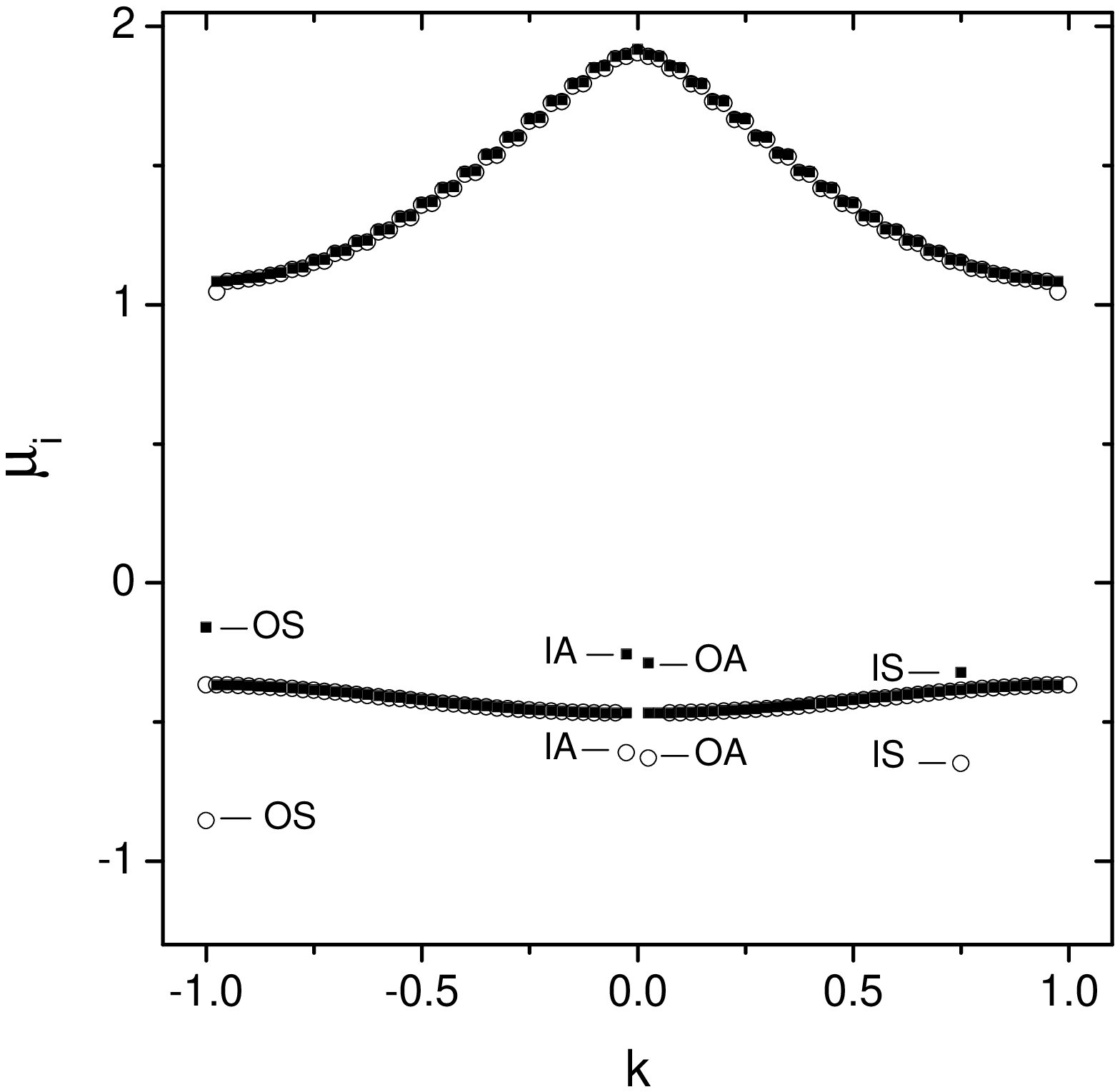}
\includegraphics[width=4.25cm,height=4.25cm,clip]{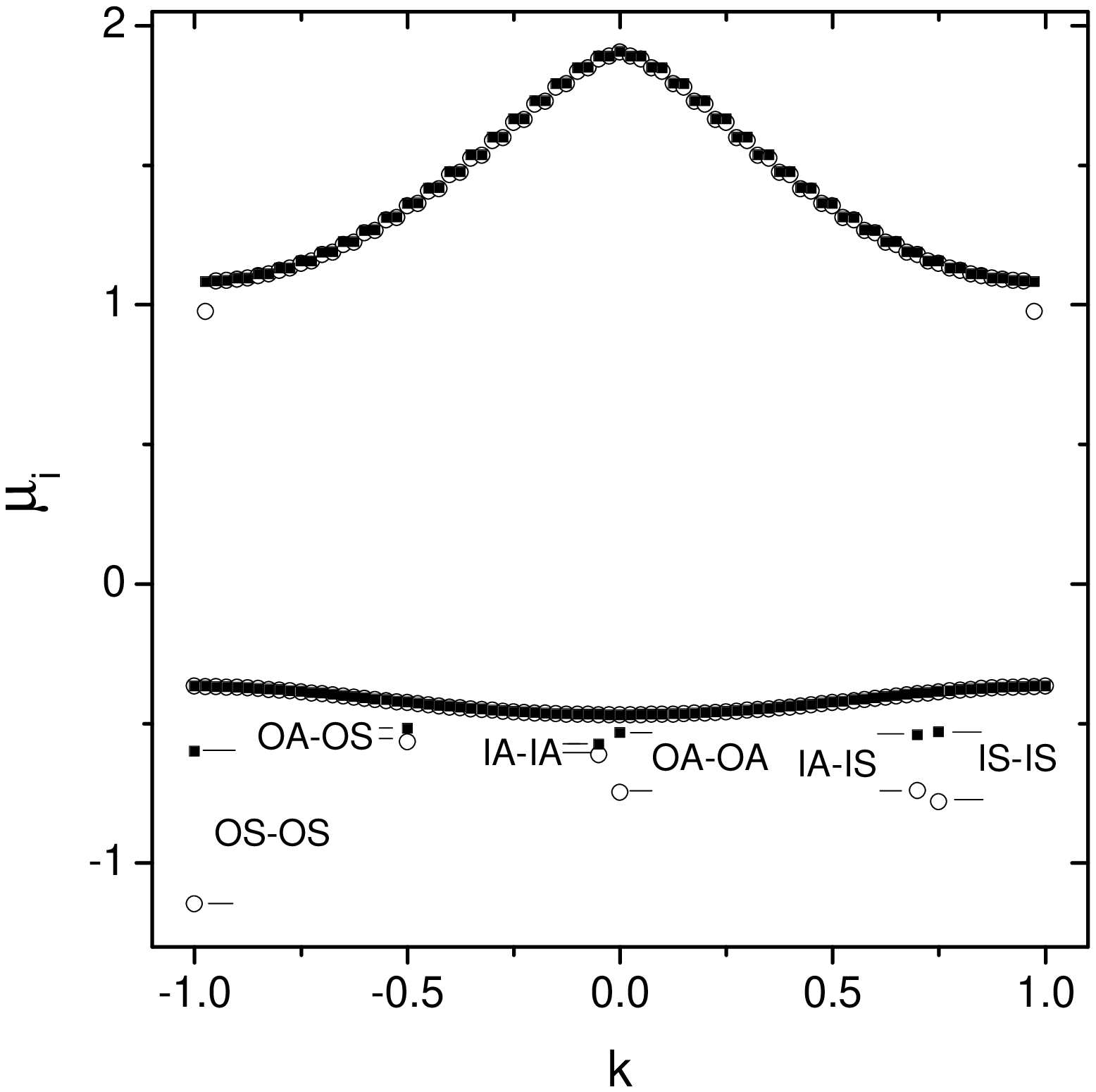}
} \caption{Band structure  and bound state levels of the two
component BEC in OL  with intra and inter species interactions
fixed as $\chi_{1}=-0.5, \chi_{2}=0.5, \chi=0$ (left panel) and as
$\chi_{1}=-0.5, \chi_{2}=0.5, \chi=-0.5$ (right panel). Other
parameters are $V_0=1.5$, $N_1=1.5, N_2=.98$.  Open circles refer
to the first component, filled squares to the second one. Notice
that in this case the equal symmetry modes are non degenerated.
Also notice the presence of localized modes near the edge of the
upper bands.} \label{fig7ms}
\end{figure}
\vskip 4cm

\begin{figure}
\centerline{
\includegraphics[width=2.65cm,height=2.65cm,angle=0,clip]{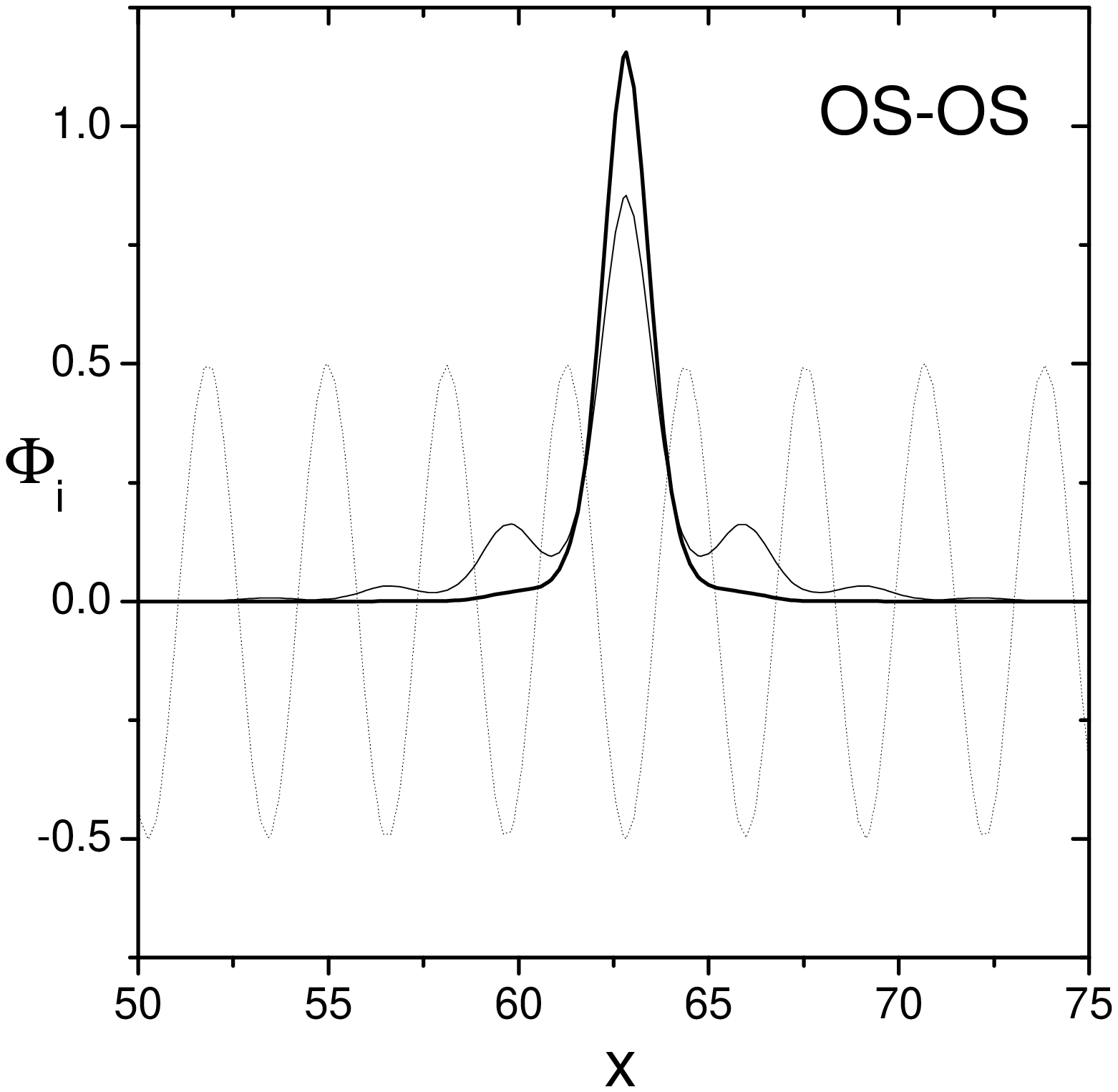}
\includegraphics[width=2.65cm,height=2.65cm,angle=0,clip]{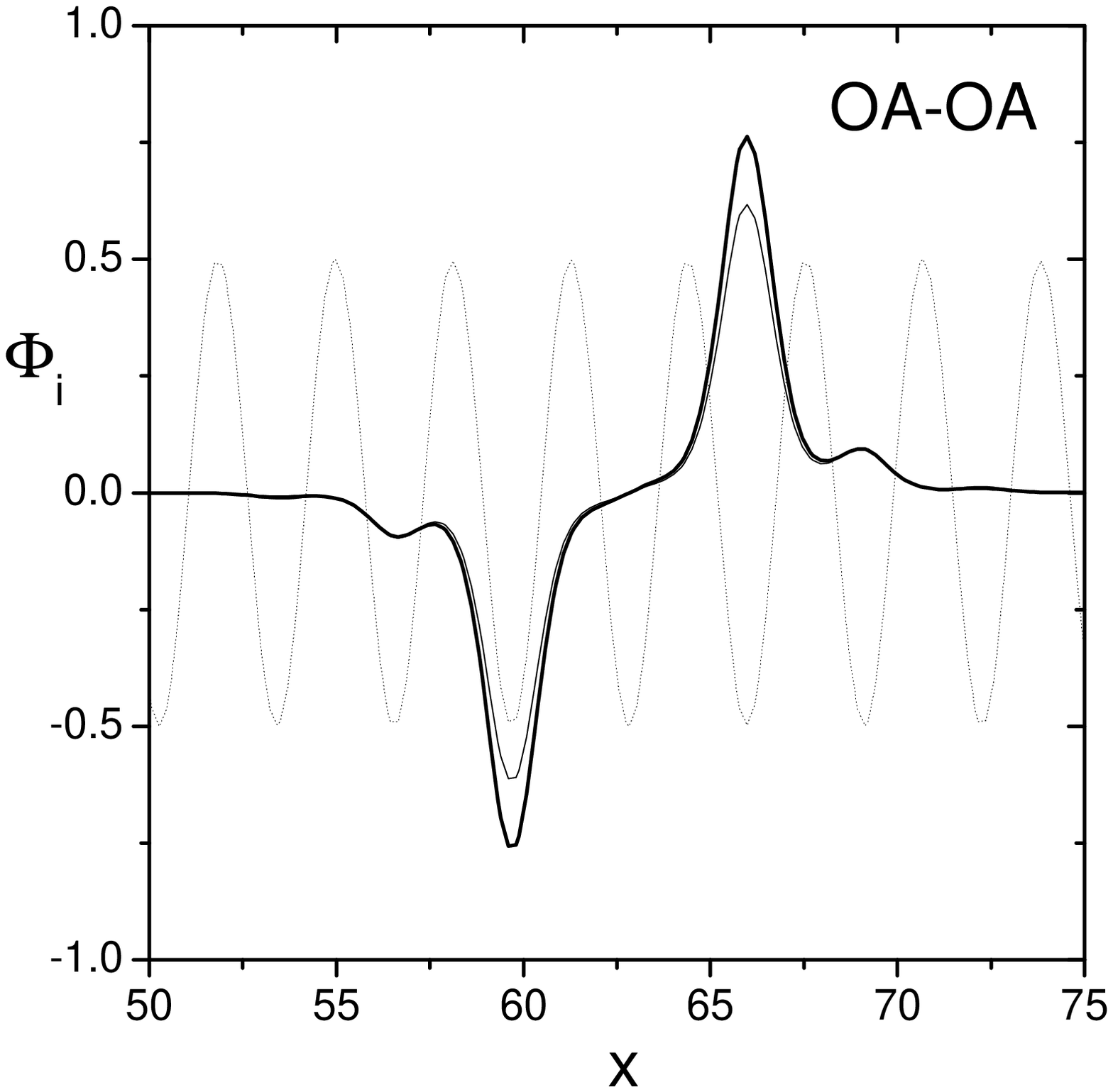}
\includegraphics[width=2.65cm,height=2.65cm,angle=0,clip]{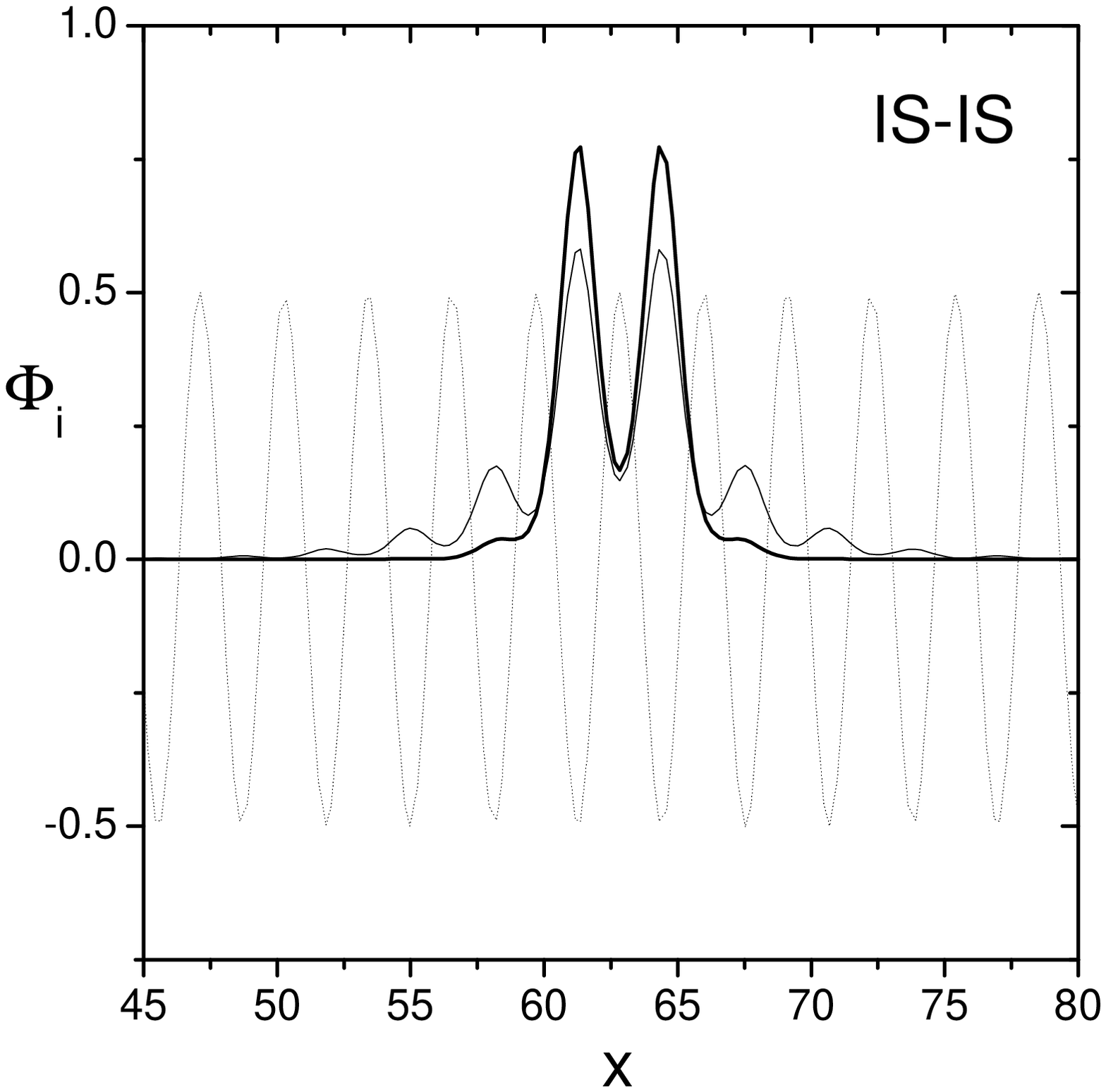}
} \centerline{
\includegraphics[width=2.65cm,height=2.65cm,angle=0,clip]{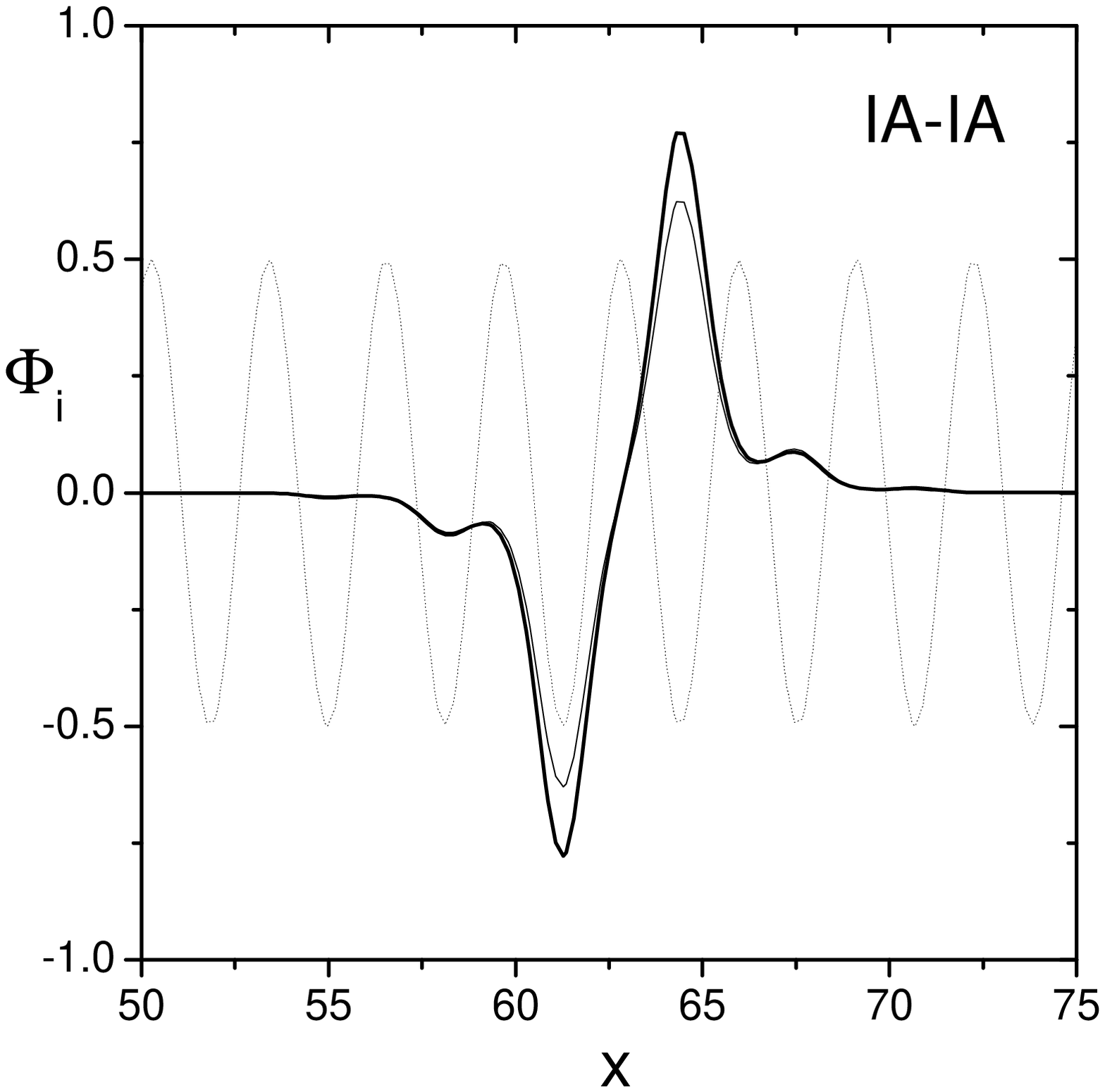}
\includegraphics[width=2.65cm,height=2.65cm,angle=0,clip]{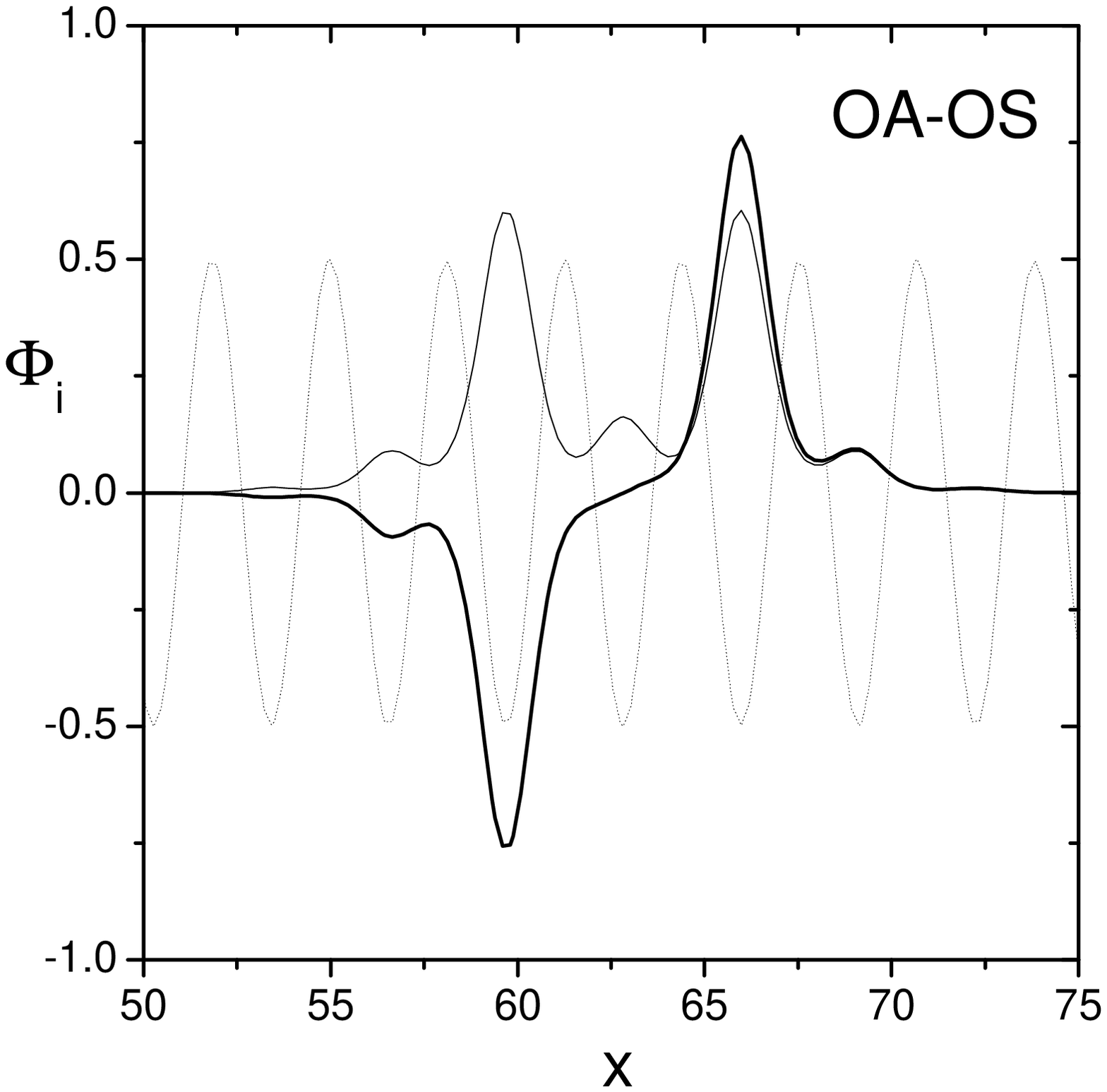}
\includegraphics[width=2.65cm,height=2.65cm,angle=0,clip]{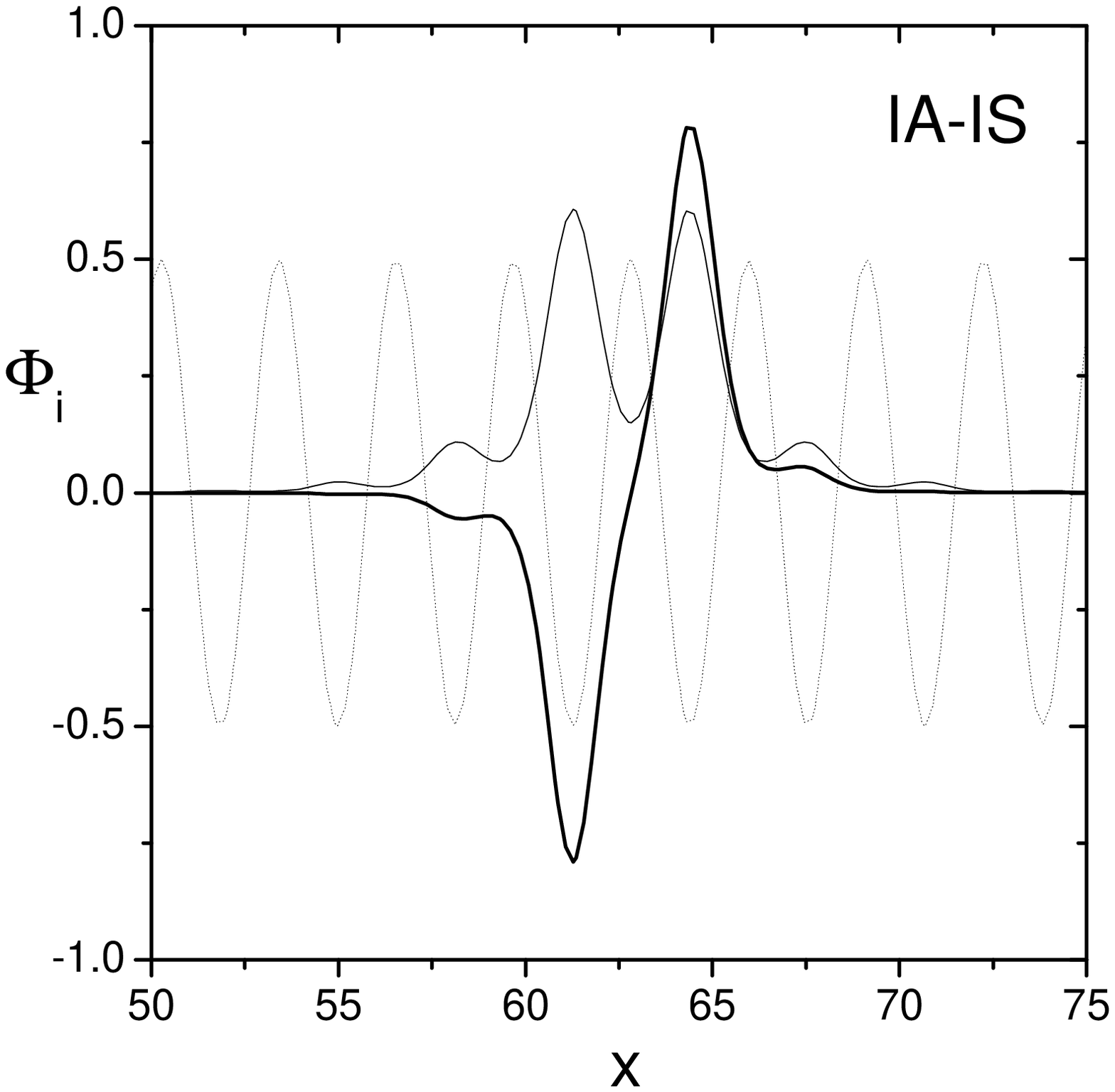}
} \caption{Condensate wavefunctions of two-component BEC in OL
corresponding to the bound state levels in the right panel of Fig.
\ref{fig7ms}. Thick and thin lines refer to first and second
component, respectively (open circles and filled squares in Fig.
\ref{fig7ms}). To identify the symmetry of the solutions, lattice
potentials (scaled by a factor 3) have been  shown by thin dotted
lines.} \label{fig8ms}
\end{figure}



We have also investigated the case of mixed interactions
$\chi_{1}<0,\chi_{2}>0,\chi<0$ (see Figs.
\ref{fig7ms},\ref{fig8ms}). In contrast with previous cases where
bound state levels were located in the semi-infinite gap or in the
first band gap, in the case of mixed interactions the bound states
are all below the first band, in spite of the fact that the second
component has a repulsive character. In absence of the
interaction, however,  the bound states of the second component
are indeed located in the first band gap as expected (see left
panel of Fig.~\ref{fig7ms}). By increasing the strength of the
interspecies interaction these eigenvalues get shifted below the
first band, this being a consequence of the attraction between the
two components (for repulsive interspecies interactions the levels
would be repelled into the gap making the formation of bound
states more difficult). Also notice that in contrast with the all
attractive and all repulsive cases, the equal symmetry localized
modes (as well as the mixed symmetry ones) are non degenerate (the
components of the bound state have different chemical potentials),
this being due to the fact that the unperturbed levels belong to
different band gaps. By changing the values of $\chi$ (by means of
Feshbach resonances) one can indeed change the position of the
bound state levels  with respect to the band structure, this
affecting the properties (in particular the stability) of the
corresponding localized state. Similarly properties are found also
for other parameter choices and for  two-component modes states of
higher band gaps.

\section{Breathers.}
\label{breathers}

The method  of construction of localized modes based on
continuation from the limit $N_2\approx 0$, can be used for
obtaining rather different solutions of the coupled GP equations
with a periodic potential, when all the nonlinearity coefficients
are equal $\chi=\chi_1=\chi_2$. Indeed, in that case the  system
(\ref{eigen}) takes the form
\begin{eqnarray}
\label{eigen_spinor_breath}
i\frac{\partial\Phi_{j}}{\partial T} = -\frac{1}{2} \frac{\partial^2 \Phi_{j}}{\partial X^2}  -V_{0}\cos(2X)\Phi_{j} \nonumber \\
+ \chi \left(|\Phi_{1j}|^2 + |\Phi_{3-j}|^2 \right)\Phi_{j}
\end{eqnarray}
and is invariant under transformation
\begin{eqnarray}
\label{invar} \varphi_1= (\Phi_1+\Phi_2)/\sqrt{2}, \quad \varphi_2
=(\Phi_1-\Phi_2)/\sqrt{2}.
\end{eqnarray}
Let us now assume
that one of the component, $\Phi_j$, is much larger than the other
one, say $N_1\gg N_2$. By using the method described in the
previous section we can find stationary solutions of the coupled
GP equations in the form $\Phi_j(X,T)=\Phi_j(X)e^{-i\mu_j T}$
(notice that the chemical potential of the large-amplitude
component is located near the top of the gap while that of the
small-amplitude component is near the bottom of the gap, such that
$\mu_1-\mu_2$ is of the order of the energy gap. Thus,
substituting the stationary solutions $\Phi_j(X,T)$ into
(\ref{invar}) we readily obtain new solutions $\varphi_j(X,T)$
characterized by the amplitudes of the same order and by {\em two}
chemical potentials, i.e. representing breathers (in the sense of
the periodically oscillating density of the condensate).

\begin{figure}[h]
\epsfig{file=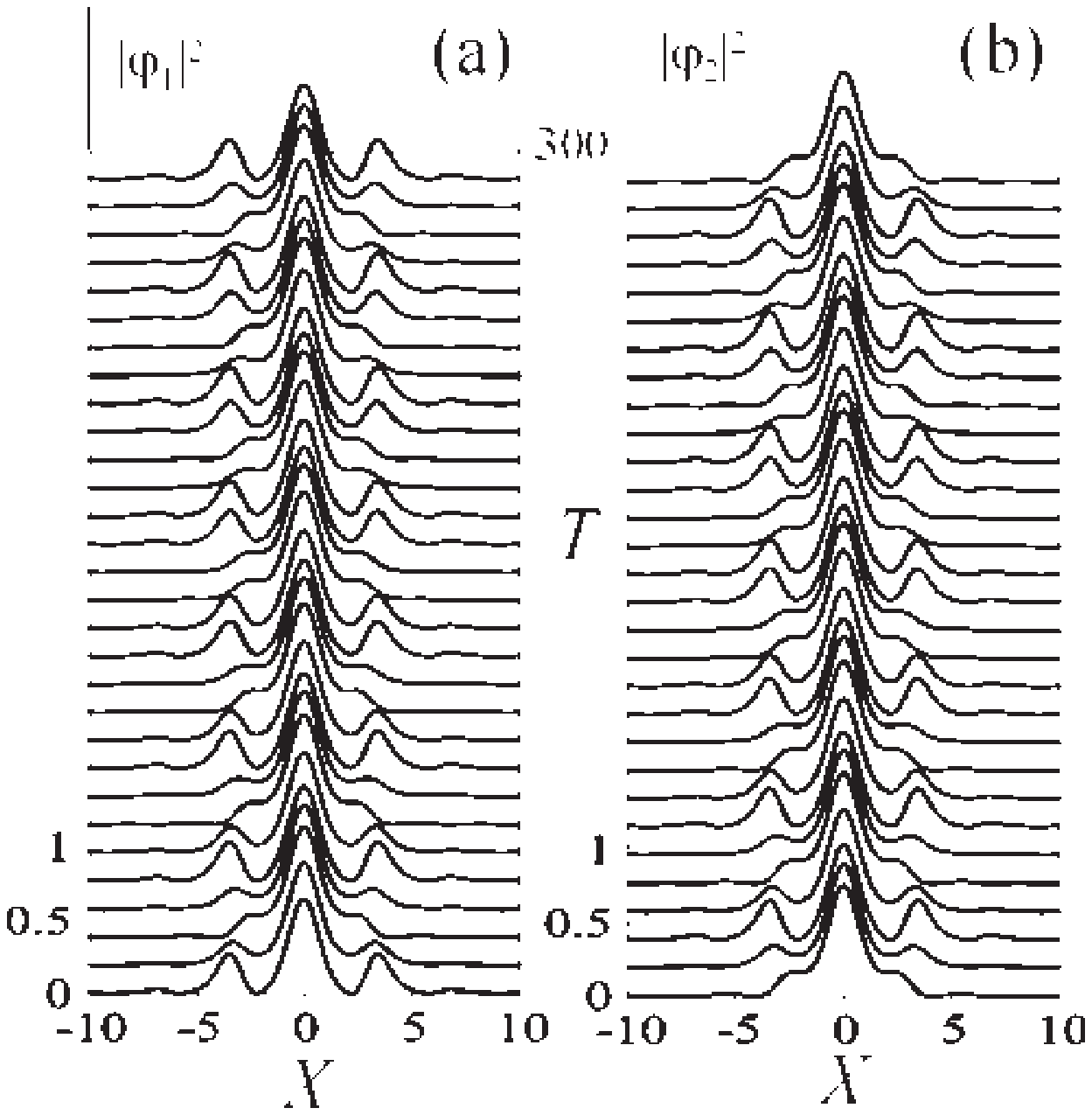,width=4cm}
\epsfig{file=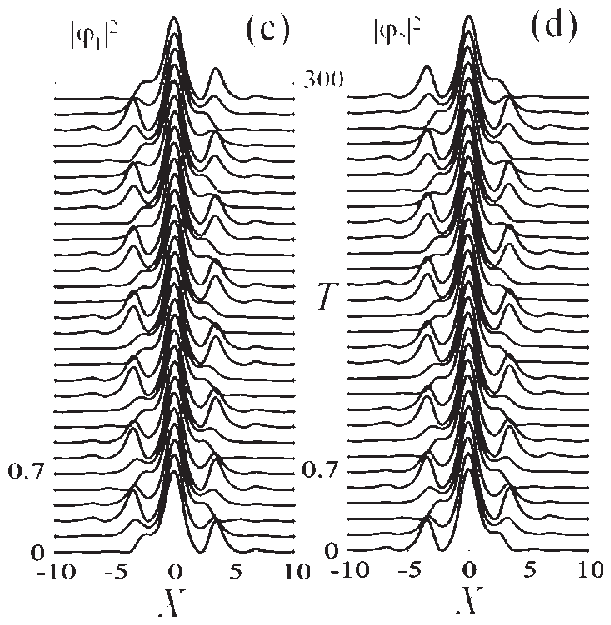,width=4cm} \caption{In [(a),(c)],
[(b),(d)] dynamics of the density of the $|\varphi_1|^2$ and
$|\varphi_1|^2$ are shown correspondingly. Initial profiles for
$\Phi_j$ are calculated for the parameters $\chi=1$, $V_0=1$,
$\mu_1=0.88$ and $\mu_2=0.1$. In [(a), (b)] the OS and [(c), (d)] the OA
profile of $\Phi_2$ was used.} \label{breath}
\end{figure}

In Fig.~\ref{breath} we show the dynamics of such breathing
solution which appears to be dynamically stable.

Besides two-components breather modes with unbalanced number of atoms,
we also find breather modes with mixed symmetry which are very stable
even for relatively large values of the number of atoms in both
components. The existence of these modes is expected both from previous
considerations and from the existence and stability of the mixed
symmetry modes with balanced number of atoms discussed in the previous
section.
%
\begin{figure}
\centerline{
\includegraphics[width=8.25cm,height=8.25cm,clip]{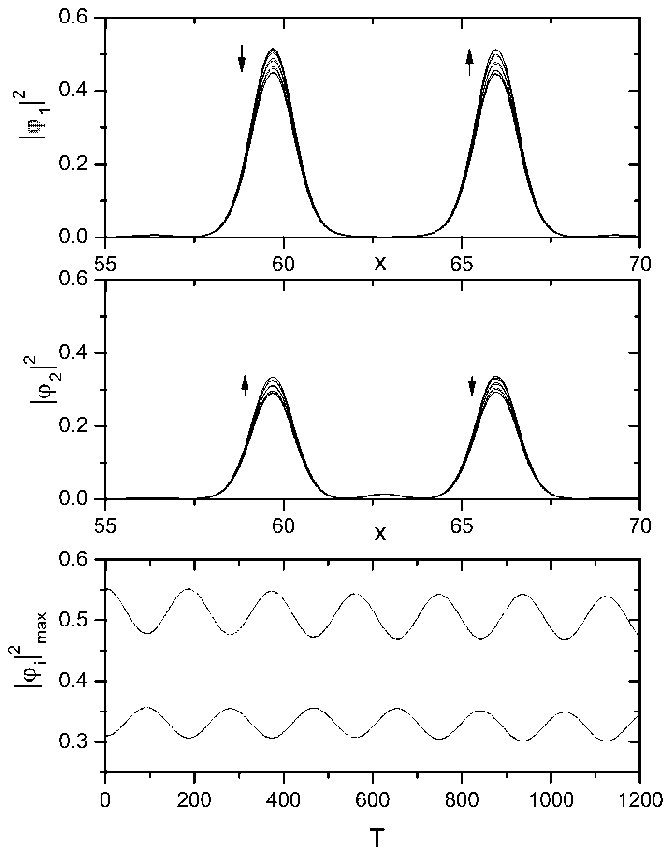}
} \caption{Gap-breather mode of OA-OS type for the case
$\chi_1=\chi_2=\chi=0.5, V_0=1.5,$ and for number of atoms
$N_1=1.5, N_2=0.98$. The top panels show ten snapshots of the
density profiles of the first (first panel) and second (second
panel) component of the gap-breather taken at regular intervals of
time during one period $T=187.86$ of oscillation. The bottom panel
show the amplitude of the left peaks of both components as a
function of time. The arrows indicate that the oscillation of the
two components occurs in opposite phase. } \label{fig6ms}
\end{figure}
\vskip 4cm
In Fig.~\ref{fig6ms} a gap-breather mode of the OA-OS type is
depicted for the case of all repulsive interactions. We see that
the oscillations of the two components are out of phase and are
very regular, persisting for very long time without any emission
of matter, this indicating that the mode is indeed very close to
an exact breather solution of the two-component GPE system.

\section{Conclusions}
\label{sec_conclusions}

In the present paper we have presented a variety of localized
modes, having stationary localized density distribution, and breathers, characterized by periodically varying density, in binary mixtures of Bose-Einstein condensates. In
particular, we have considered modes having either close or very different
numbers of atoms (we termed them balanced and unbalanced modes) and characterized by the same or different symmetries of the atomic distributions in each components. We have suggested algorithms of systematic finding families of the modes. We exploited the fact that unbalanced modes bifurcate from the linear defect modes, the defect being understood as a localized excitation of one of the components in the absence of another component. This allowed us construct exact breather modes in a systematic way. 

We have introduced the classification of the modes on the basis of
four possible distributions of each of the components, what  in the general case results in  the modes
of 8 different types, which can be designated as BC$_1$-BC$_2$,
where B stands either for "O" in the case of the distribution
centered in a local minimum of the periodic potential or for "I"
in the case of the mode centered in the maximum of the potential,
and C$_j$ stands for S or A depending whether the wavefunction of
the the $j$-th component is even (S) or odd (A).

Although our analysis was concentrated on spinor condensates, it allows for straightforward generalization for arbitrary two-component system described by coupled nonlinear Schr\"odinger equations with periodic coefficients.

\acknowledgments

HAC acknowledges support of the FCT thorough the grant SFRH/BD/23283/2005.
VAB was supported by the FCT grant SFRH/BPD/5632/2001. VVK acknowledges
support from Ministerio de Educaci\'on y Ciencia (MEC, Spain)
under the grant SAB2005-0195. M. S. acknowledges partial financial
support from the MIUR, through the inter-university project
PRIN-2005, and from the (CNISM) Consorzio Nazionale
Interuniversitario per le Scienze Fisiche della Materia. The work
of HAC, VAB, VVK, and GLA was supported by the FCT and FEDER under the
grant POCI/FIS/56237/2004.

\appendix

\section{Some comments on the shooting method}
\label{app:1}

In this Appendix we briefly outline the implementation of the shooting
method for finding localized solutions of (\ref{eigen}). Let
$\mu_{1,2}$ and $V_{01}$, $V_{02}$ be fixed and $\mu_{i}$ belong
to gaps of the potentials $V_{0i}U(X)$. In the limit $X\to\infty$
the equations in (\ref{eigen}) separate. Both of them become
linear Hill equations and the asymptotics of $\Phi_{1,2}(X)$ as
$X\to\infty$ are given by the formula ($j=1,2$)
\begin{eqnarray}
\Phi_j(X)\sim C_j e^{-\gamma_1 X}H_j(X)\,,
\label{AsCond}
\end{eqnarray}
where $2\pi$-periodic functions $H_{1,2}$ and values
$\gamma_{1,2}>0$ are uniquely defined by the parameters of
equations. The parameters $C_{1,2}$ specify the solutions of
(\ref{eigen}) which tend to zero at $X\to\infty$. If $C_{1,2}$ are
fixed the solution $\Phi_{1,2}(X;C_1,C_2)$, subject to the
conditions (\ref{AsCond}) at $X\to\infty$, can be
found by shooting backward starting from large enough value of $X$
where this asymptotics is valid. Let us consider, for the sake of
definiteness, {\it even} localized solutions of (\ref{eigen}).
They have to satisfy the conditions
\begin{eqnarray}
\frac{d\Phi_1(0;C_1,C_2)}{dX}=0;\quad\frac{d\Phi_2(0;C_1,C_2)}{dX}=0.
\label{SysC1C2} 
\end{eqnarray} This is the system of two
equations for two variables $C_1$ and $C_2$. Generically, the
solutions of this system are {\it isolated} [in the sense that if
$C_1^0$ and $C_2^0$ is a solution there is some neighborhood of
$(C_1^0;C_2^0)$ which does not contain other solutions of
(\ref{eigen})]. Similar arguments hold for odd and also for
non-symmetric solutions of (\ref{eigen}). In this way one arrives
at the conclusion formulated in the text.

\end{document}